\documentclass[11pt,letterpaper,twocolumn,english, double-column]{article}
\usepackage{amsmath}
\usepackage{mathptmx}
\usepackage{newtxmath}

\usepackage[T1]{fontenc}
\usepackage[utf8]{inputenc}
\usepackage{fancyhdr}
\usepackage{babel}
\usepackage{array}
\usepackage{booktabs}
\usepackage{graphicx}
\usepackage{wasysym}
\usepackage[authoryear,numbers]{natbib}
\PassOptionsToPackage{normalem}{ulem}
\usepackage{ulem}
\usepackage[unicode=true]
 {hyperref}
\usepackage{xcolor}
\usepackage{gensymb}
\usepackage{bm}
\usepackage{threeparttable}
\usepackage{subfig}
\usepackage{placeins} 
\usepackage{enumitem}
\usepackage{appendix}
\usepackage{stfloats}

\makeatletter

\providecommand{\tabularnewline}{\\}

\input glyphtounicode
\usepackage{graphicx}
\usepackage{dcolumn}
\usepackage{caption}
\usepackage{graphicx}
\usepackage{titlesec} 

\usepackage[T1]{fontenc}
\usepackage{ae,aecompl}

\usepackage{fancyhdr}

\usepackage{etoolbox} 
\usepackage{blindtext}  %only for dummy text, not for use in documents
\usepackage{enumitem}

\titleformat{\section}{\normalfont\bfseries\fontsize{14pt}{16pt}\selectfont}{\thesection.}{0.5em}{}
\titleformat{\subsection}{\normalfont\bfseries\fontsize{12pt}{14pt}\selectfont\itshape}{\thesubsection.}{0.5em}{}
\titlespacing\section{0pt}{12pt plus 4pt minus 2pt}{0pt plus 2pt minus 2pt}
\titlespacing\subsection{0pt}{6pt plus 4pt minus 2pt}{-6pt plus 2pt minus 2pt}
\titlespacing\subsubsection{0pt}{6pt plus 4pt minus 2pt}{-6pt plus 2pt minus 2pt}

\renewcommand{\@listI}{%
\leftmargin=25pt
\rightmargin=0pt
\labelsep=5pt
\labelwidth=20pt
\itemindent=0pt
\listparindent=0pt
\topsep=3pt plus 2pt minus 4pt
\partopsep=3pt plus 1pt minus 1pt
\parsep=3pt plus 1pt
\itemsep=\parsep}

\title{Modeling Utah FORGE 2022 EGS Hydraulic Stimulations: Tensile Hydraulic Fractures versus Fluid-Induced Dilatant Shear Ruptures}
\author{Sylvain Brisson and Brice Lecampion\\
\textit{Geo-Energy Laboratory, Ecole Polytechnique Fédérale de Lausanne, Switzerland}}

\begin{document}
 
\twocolumn[
  \begin{@twocolumnfalse}
    \maketitle
    \begin{abstract}

We investigate two hydraulic stimulation stages performed in April 2022 at the Utah FORGE enhanced geothermal system test site using analytical and numerical models for tensile hydraulic fractures and fluid-induced dilatant shear fractures. The two injection stages differ primarily by the viscosity of the fracturing fluid: low-viscosity slick-water for the first stage and high-viscosity cross-linked gel for the second. Despite similar injection rate schedules and well-head pressure responses, the two stages exhibit markedly different post–shut-in microseismic behavior. The cross-linked gel stage shows sustained microseismic activity for several hours after shut-in, whereas the slickwater stage exhibits an immediate decrease.
For the cross-linked gel stage, the located microseismic events benefit from adequate seismic acquisition coverage. They reveal the development of a planar radial fracture and allow confident retrieval of the fracture extent evolution with time. We demonstrate that this evolution follows the scalings predicted for viscosity–storage–dominated radial hydraulic fracture by analytical models, both during injection and post–shut-in, providing strong evidence for the development of a planar tensile hydraulic fracture. We further show that leak-off is required to reproduce the final fracture extent. In contrast, the immediate arrest observed during the slick-water stage suggests either a transition to a toughness- or leak-off–dominated hydraulic fracture regime, or the development of a fluid-induced shear fracture. Using scalings from analytical models of fluid-induced shear rupture, we show that the slickwater stage could plausibly correspond to a dilatant shear fracture, provided sufficient dilatancy, whereas this hypothesis is invalidated for the cross-linked gel stage.
We confirm these insights, derived from analytical models, using a 3D axisymmetric fully-coupled hydro-mechanical numerical model capable of resolving both tensile and shear failure modes, and including leak-off. Finally, we propagate uncertainties in the in-situ stress state and natural fracture orientations through this numerical model to assess their impact on injection pressures.

\end{abstract}

\subsection*{Highlights}

\vspace{0.5cm} 

\setlist{itemsep=0pt}

\begin{itemize}
\item Fracture extent evolution with time recovered from localized micro-seismic events
\item Post-shut-in growth/arrest behavior explained by  linear  hydraulic fracture mechanics
\item Observed fracture growth matches scalings from viscosity-storage dominated radial hydraulic fracture
\item Hydraulic stimulation stages modeled as tensile or fluid-induced shear fractures using a fully-coupled hydro-mechanical solver
\item Hydro-shearing hypothesis ruled out for cross-linked gel stage but remains plausible for slick-water stage
\end{itemize}

\vspace{1cm} 

\end{@twocolumnfalse}             
]

% ----------------------------------
% INTRODUCTION
% ----------------------------------
\section{Introduction}

In the context of energy transition away from fossil fuels, deep geothermal energy can provide decarbonized, non-intermittent energy \citep{Ricks2022}. Today, however, it is limited to favorable situations of strong geothermal gradients coupled with natural permeability of the geothermal reservoir. Enhanced geothermal system (EGS) technology promises to increase its potential by allowing heat extraction from hot but not naturally permeable media by artificially increasing this permeability \citep{Olasolo2016}. This can be done via hydraulic stimulation: injecting fluid at high pressures to mechanically shear or open fractures. We make the distinction between EGS projects that have relied on the dilatant shearing of pre-existing stressed fractures, a technique commonly referred to as hydro-shearing, and EGS project that have aimed for hydraulic fracturing (HF) : the opening of new tensile fractures. The Soultz-la-forêt project \citep{Evans2005} can be cited as an example of a hydro-shearing-based EGS project. The more recent Utah FORGE EGS test site \citep{moore2019utah} or the neighboring Fervo project \citep{Norbeck2023} have relied more on HF. Hydraulic fracturing is routinely used for other purposes, notably the stimulation of oil and gas reservoirs.
Hydraulic fracture growth is a quasi-static/stable process controlled by the volume of fluid injected. It is always associated with microseismicity: small-scale dynamic instabilities due to heterogeneities of stress or mechanical properties. 
Hydro-shearing consists of fluid-induced frictional ruptures that propagate at a pressure lower than the minimum confining stress. For such frictional ruptures, a dynamic run-away event can occur if the residual frictional strength is lower than the ambient shear stress \citep{GaGe12,SaLe24}. Such large-scale dynamic run-aways can cause hazardous large seismic events. Hydraulic stimulations for EGS take place at great depths, usually between 2 and 5 km, in complex geological media from which we have limited knowledge. It is therefore challenging to predict the outcome of a hydraulic stimulation in terms of permeability enhancement and induced seismicity \citep{McClure2014}. This has led to the failure of EGS projects due to the inability to achieve economical flow rates (Habanero EGS project \citep{Hogarth2017} among others) or the occurrence of large induced seismic events (induced event of magnitude 3.4 in the Basel EGS project \cite{Deichmann2009}).

EGS projects relying on hydro-shearing have been effectively modeled as the propagation of fluid-induced aseismic shear ruptures on preexisting stressed faults (see \cite{Saez2023} for an example of modeling of a stimulation performed at the Soultz-la-forêt test site). 

For EGS projects relying on HF, we benefit from the work done on the modeling of planar radial hydraulic fractures \citep{Savitski2002,Detournay2016}. The planar radial HF model is well-suited for modeling HF in crystalline rocks that can be considered isotropic (regarding elasticity and fracture toughness) and relatively uniform. 
% This is a key difference with the modeling of HF that develop in layered media in which oil and gas reservoirs develop.

The Utah FORGE (Frontier Observatory for Research in Geothermal Energy) project is an EGS test site that has been under development since 2019. It is located in the Basin and Range formation in Utah \citep{moore2019utah}. It aims to demonstrate the feasibility of harnessing the heat of deep naturally impermeable crystalline rocks under strict risk management procedures, especially regarding induced seismicity hazard. As the project relies on multi-stage hydraulic fracturing of cased deviated sub-horizontal wells, with the use of proppant to maintain fracture conductivities, the stimulation procedures are similar to what is done for oil and gas hydraulic fracturing \citep{UtahPhase3B}. The reservoir is planned to be developed in a hard granitoid rock with high in-situ temperatures of about 200$^\circ$C at 2500 m.

We propose using analytical and numerical models of hydraulic fracture and hydro-shearing to gain insights into hydraulic stimulations performed in 2022 at the Utah FORGE EGS project. 

We rely on a detailed analysis of the located microseismic events to derive the evolution of the fracture size with time. This fracture size derived from located microseismicity is a better observation than the measured well-head pressure, which is strongly affected by physical processes that are local to the so-called fracture-entry region (which we do not intend to model here). We study this fracture size evolution using planar radial HF and hydro-shearing models (for which scalings and analytical solutions are known from previous works), to gain insights on the stimulations. We then confirm these insights using a fully-coupled hydro-mechanical 3D axisymmetric solver, suited for both HF and hydro-shearing modeling. The low numerical cost of this solver enables us to perform uncertainty propagation of the in-situ stress state and natural fractures orientation.

The same stimulations have been modeled in previous studies, but using different modeling frameworks and under different objectives. In \cite{Xing2021}, the authors rely on a 3D lattice-based solver, XSite, to model the stimulations a priori either as pure hydraulic fractures or as the opening/shearing of discrete fracture network (DFN) realizations. As this study was performed before the actual stimulations took place, it did not aim to analyze the observations gathered during the stimulations. In \cite{McClure2023}, the authors model one of the stimulation stages as a hydraulic fracture. Like us, they rely on a planar hydraulic fracture model motivated by the planarity of the observed induced microseismicity. The model they use incorporates additional processes and physics we do not account for: notably, proppant transport, thermal energy transport, and stress gradients. They also rely on empirical laws such as scale-dependent toughness and leak-off, deviations from the cubic law due to roughness, and decreasing near-wellbore tortuosity due to erosion by proppant. We demonstrate that our simpler modeling approach, supported by well-understood assumptions, allows us to recover the first-order behavior of the stimulations.

The paper is organized as follows: Section 2 presents the field data, including rock properties, stress state, natural fractures, and microseismic observations; Section 3 describes the analytical and numerical modeling approaches; and Section 4 presents the modeling results under both hypotheses, along with uncertainty propagation of the in-situ stress state.

% ----------------------------------
% DATA
% ----------------------------------
\section{Field data from FORGE 2022 hydraulic stimulations}

Three stimulation stages have been performed in 2022 in the sub-horizontal terminal part of the deviated well 16A at a depth of around 2500 m (true vertical depth). The reservoir is a competent unweathered granitoid rock It is naturally fractured but with non-conductive fractures, which makes the overall rock-mass relatively impermeable. The upscaled permeability, taking into account the presence of natural fractures, has been estimated to be  below $1.5.10^{-16}$ m$^2$ (99\% CI) in \cite{Finnila2023}. The stimulation stages were performed one after the other, successfully using bridge plugs to isolate them. One was performed in the open-hole (without casing) toe section of the well, which we will not study here. The two other stages, which will be studied here, were performed behind casing with perforated sections spanning over 6 meters. They differ primarily by the viscosity of the fluid used to perform the stimulation. In stage 2 (out of the 3 performed), water with friction reducer additives has been used. The fluid obtained, referred to as slick-water, has the same viscosity of water. The later is given at $1.28.10^{-4}$ Pa.s at the reservoir conditions (225$^\circ$C and 20 MPa) by the IAPWS (\cite{Wagner2002}). We will refer to this stage as the "slick-water" stage. In the third stage, a polymer (carboxymethyl hydroxypropyl guar gum, or CMHPG) along with a cross-linking agent, and other additived, was added to water to obtain a high-viscosity gel. It will be referred to as the "cross-linked gel" stage. The initial viscosity of this gel is measured to be about 1 Pa.s. However, at the high temperature conditions encountered at the injection depth, the polymer likely degraded over the time scale of the stimulation, leading to a significant drop in fluid viscosity. This change in viscosity was evaluated using a rheometer under high temperature conditions in \citep{moore2019utah}: the viscosity decreased from about 1 Pa.s initially to about 0.01 Pa.s over a time span of 2 hours when exposed to temperatures close to the one encountered in the reservoir.
The associated data is shown in Appendix \ref{Appendix:fcrosslinked_gel_viscosity}. 
In our modeling, we assume the fracturing fluid to be Newtonian and of constant viscosity; we will settle for a base value of 0.1 Pa.s (in line with the values chosen in previous works \citep{McClure2023, Xing2021}), along with a sensitivity analysis to establish how this choice affects our results.
% We show that under this simplification we still recover the first order behavior of the stimulation.

% TODO: improve this part

\subsection{Rock mass mechanical properties}
\label{section:data:rock_mass_properties}

Our modeling requires the elastic moduli and the mode I fracture toughness of the rock. These have been measured in the lab on intact core samples retrieved during drilling operations; the values are reported in \cite{McLennan2018}. They obtain a Young's modulus of $E = 47.5 \pm 11.5$ GPa, a Poisson's ratio of $\nu = 0.27 \pm 0.7$ and a mode I fracture toughness of $K_{Ic} = 1.8 \pm 0.4$ MPa.m$^{1/2}$. To account for the fact that the elastic moduli of the naturally fractured rock mass are lower than the value measured on intact core samples, we apply to the Young modulus the formula of \cite{Hoek2006}. Using a disturbance factor $D=0$ (all joints are closed) and a geological strength index of $GSI = 80$ (lower bound for a very fresh/unweathered
rock mass), we obtain a value of $E_{rm} = 42$ GPa for the Young modulus of the rock mass. This 10\% decrease of elastic moduli does not have a first-order effect on our modeling outcome. For the mode I fracture toughness, it is expected for the large-scale observed value to be larger than the lab-measured value. In \cite{McClure2023}, the authors model the cross-linked gel stage as a hydraulic fracture and find a value of 4$\times$ the lab-measured value to be appropriate for $K_{Ic}$. In \cite{Xing2021} the authors opt for a value of $K_{Ic} =3$ MPa.m$^{1/2}$. We settle for a value of $K_{Ic} =3.6$ MPa.m$^{1/2}$, twice the lab-measured value.

\subsection{In-situ state of stress}
\label{section:data:state_stress}

\begin{figure}[h!]
    \centering
    \includegraphics[width=0.5\textwidth]{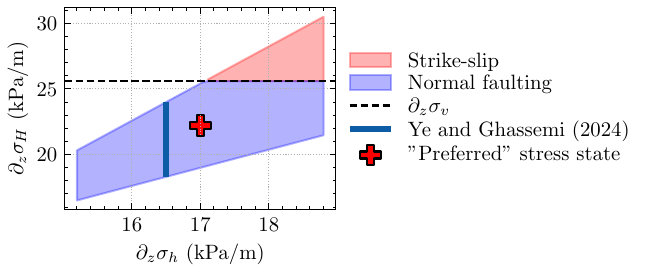}
    \caption{In-situ state of stress model obtained by combining results of \cite{moore2019utah}, \cite{Xing2020}, \cite{Ye2024} and \cite{Kelley2024}}
    \label{fig:state_of_stress_model}
\end{figure}

Several studies have aimed at characterizing the in-situ state of stress of the Utah FORGE reservoir. Most of these studies assume that one of the principal stresses is vertical. Under this assumption, the state of stress is characterized by the magnitudes of its 3 principal components, the in-situ pore pressure, and the orientation of its horizontal components. In the following, we use $\sigma_v$, $\sigma_H$, and $\sigma_h$ to refer to the vertical, maximum horizontal, and minimum horizontal stresses, respectively. Their 3 respective magnitudes are assumed to vary linearly with depth within the reservoir. The corresponding vertical gradients are denoted by $\partial_z\sigma_v$, $\partial_z\sigma_H$, and $\partial_z\sigma_h$. We use $\alpha_H$ to refer to the azimuth (anti-clockwise angle with respect to true North) of the maximum horizontal stress. The magnitude of $\sigma_v$ is obtained by integration of the density log. It is well constrained, and it is reported in \cite{Moore2019} to be of $\partial_z\sigma_v \simeq 25.6$ kPa/m in the region of interest. The magnitude of $\sigma_h$ is less well constrained and is evaluated by picking the closure pressure in small hydraulic fracturing tests. This has been done within the same formation in a vertical well neighboring the well 16A by \cite{Xing2020}. The authors find that when performing multiple micro-hydraulic fracturing tests within the same zones , the obtained values of $\partial_z\sigma_h$ span about uniformly the interval $15.2$ kPa/m to $18.8$ kPa/m. We take as our "preferred" value the center of this interval; $\partial_z\sigma_h\simeq 17$ kPa/m. The magnitude of $\partial_z\sigma_H$ is even more poorly constrained than $\partial_z\sigma_h$. Work has been done in \cite{Ye2024} and \cite{Kelley2024} to model it using the orientation of observed drilling-induced tensile fractures in wells 16A and 16B, respectively. In \cite{Ye2024} (which uses data of the well of interest for us) the authors assume $\partial_z\sigma_h = 16.5$ kPa/m and obtain $\partial_z\sigma_H \in [18.3, 24]$ kPa/m. In \cite{Kelley2024}, the authors perform a sensitivity analysis on how the choice of $\partial_z\sigma_h$ affects the outcome on $\partial_z\sigma_H$, they observe a strong linear variation of the upper and lower bounds on $\partial_z\sigma_H$ depending on $\partial_z\sigma_h$: $d(\partial_z\sigma_H^+)/d(\partial_z\sigma_h) \simeq 2.8$ and $d(\partial_z\sigma_H^-)/d(\partial_z\sigma_h) \simeq 1.4$. 
We use all these reported values to build a probabilistic stress state model. The vertical stress gradient is taken as granted, taking the value introduced above.  The in-situ pore pressure is taken as hydrostatic from the surface: $\partial_z p_0 \simeq 9.3$ kPa/m. For $\partial_z\sigma_h$ we take is uniform over for the bounds reported by \cite{Xing2020}. For $\sigma_H$ we take the bounds on $\partial_z\sigma_H$ obtained by \cite{Ye2024}, assuming a single value of $\partial_z\sigma_h$, and we extend it over the span $\partial_z\sigma_h$ by using the results of the sensitivity analysis of \cite{Kelley2024}. Following that choice, the magnitudes of the horizontal stresses are taken as non-independent random variables. They span uniformly the space shown in Fig. \ref{fig:state_of_stress_model}. We observe that under these hypotheses, both a normal-faulting regime and a strike-slip regime are possible, although a normal-faulting regime is more likely. We constrain the orientation of the horizontal stresses $\alpha_H$ by looking at the orientation of the drilling-induced tensile fractures. The observed distribution of these orientations is well fitted by a normal distribution centered on N20$^\circ$E with a standard deviation of 10$^\circ$ while truncated between N5$^\circ$E and N30$^\circ$E.

\subsection{Natural fractures}
\label{section:data:natural_fractures}

\begin{figure}[h!]
    \centering
    \includegraphics[width=0.5\textwidth]{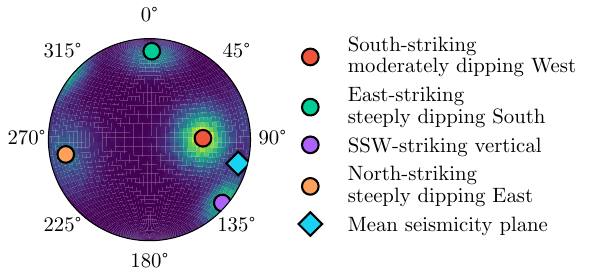}
    \caption{Stereonet representation of the 2D histogram of the orientation of the 130'000 fractures contained in one of the realizations of the DFN model of \cite{Finnila2023}, along with the mean orientation of the 4 fracture sets identified by the authors. The mean plane of the located induced microseismicity (see section \ref{section:data:seismic_radius}) is also reported.}
    \label{fig:dfn_model_orienttaion}
\end{figure}

The reservoir is naturally fractured, but these fractures are poorly hydraulically conductive. We rely on the work of \cite{Finnila2023} where the authors have developed a probabilistic discrete fracture network (DFN) model of the FORGE reservoir using fractures observation from borehole logs and outcrops. Their work identifies four distinct fracture sets. They give access to realizations of their probabilistic DFN model containing on the order of 100'000 fractures. The orientation of the fractures of such realizations is shown in Fig. \ref{fig:dfn_model_orienttaion}.

Our modeling also requires the mechanical properties under shearing of such natural fractures. Triaxial direct-shear tests on samples from the Utah FORGE site containing natural fractures have been conducted by \cite{Iyare2025}. They measure residual friction angles between $32^\circ$ and $48^\circ$ (friction coefficient between $0.51$ and $0.7$), dilation angles between $3^\circ$ and $7^\circ$ (dilation coefficient between $0.05$ and $0.12$), and cohesion of about $3$ MPa.

\subsection{Injection rate, wellhead pressure, and microseismic emissions during stimulations}
\label{section:data:stimulation_data}

\begin{figure}[ht!]
    \centering
    \includegraphics[width=1\linewidth]{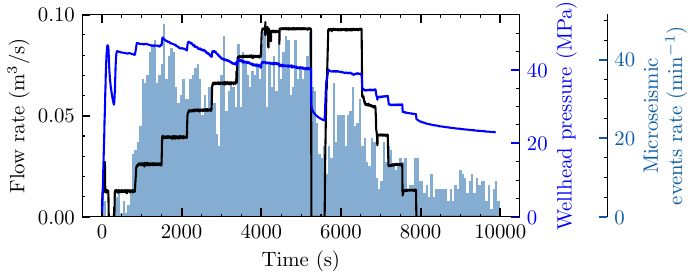}
    \caption{Stage 2 (slick-water): injection flow-rate, wellhead pressure, and microseimic emissions rate.}
    \label{fig:inj_stage2}
\end{figure}

\begin{figure}[ht!]
    \centering
    \includegraphics[width=1\linewidth]{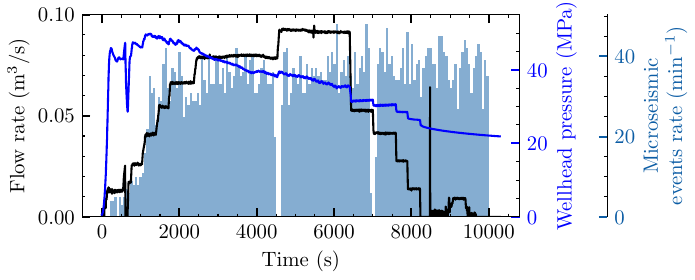}
    \caption{Stage 3 (cross-linked gel): injection flow-rate, wellhead pressure, and microseismicity rate.}
    \label{fig:inj_stage3}
\end{figure}

The two stimulation stages were conducted by controlling the injection flow rate at the wellbore head. Both stages followed a similar injection rate schedule as shown in figures \ref{fig:inj_stage2} and \ref{fig:inj_stage3}, It consists in a stepwise increase to a plateau, followed by a stepwise descent, over a 3 to 4 hours duration. The maximum injection rates and total injected volumes are similar for both stages; around 100 L/s for both stages for the rate, 418 m$^3$ for stage 2, and 507 m$^3$ for stage 3 for the total injected volume. The observations available are the wellhead pressure, the fluid injection rate, and the induced microseismicity recording (via three-component geophones located in neighboring vertical wells). This microseismicity data has been processed by \cite{GES2022} to build a catalog of microseismic events induced by the different stimulations.

% During stage 2, a temporary arrest in the injection was performed within the highest rate plateau to investigate how the pressure and the induced microseismicity would respond.

During both stages, the wellhead pressure responses were similar, as shown in figures \ref{fig:inj_stage2} and \ref{fig:inj_stage3}. During the first 5 minutes of injection, we observe the linear pressurization of the fluid in the well, until breakdown occurs between 40 MPa and 50 MPa. At this pressure, a fracture is opened or sheared, which leads to a strong increase in its hydraulic conductivity and hence a decrease in the wellhead pressure. As the injection continues and the flow-rate increases, the wellhead pressure continues to decrease, indicating the effective stimulation of the reservoir.

We estimate the wellbore storage by fitting a line to the initial linear pressurization observed in stage 2. We obtain a value of $C = 2.8 10^{-8}$ Pa/m$^3$. This value is on par with the product of the well volume with the compressibility of water.

% We estimate the wellbore storage and the initial transmissibility of the rock near the wellbore using the pressure evolution during the first 300 seconds of the stage 2, which consist in a step increase of the rate up to 12 L/s maintained during about 150 seconds followed by a about 150 seconds of shut-in.  The wellbore storage combines the compressibility of the well and of the fluid with-in. We estimate it at $C = 2.8 10^{-8}$ Pa/m$^3$ by fitting a line to the initial pressurization. This value corresponds roughly to the product of the well volume with water compressibility.

% The initial transmissibility is estimated using the so-called radial flow model, which supposes a line-injection in an infinite reservoir of width $w_h$ and homogeneous permeability $k$. We obtain a value of the initial transmissibility of $[kw_h]_0 = 1.3 10^{-13}$ m$^3$.

The microseismicity rate throughout the stimulations is also shown in  Fig. \ref{fig:inj_stage2} and \ref{fig:inj_stage3}. Both stages exhibit a similar behavior during the stepwise increase of the injection rate, with the microseismicity quickly building up before reaching a plateau at about 50 events/min. However, the microseismicity rates are significantly different between the two stages during the step-wise decrease of the injection rate. During the slick-water stage, the microseismicity rate immediately slows down as the injection rate is reduced, with limited residual microseismic activity after the end of the injection (shut-in). On the contrary, during the cross-linked gel stage, the microseismicity rate appears unaffected by the decrease in injection rate. This observation tends to show that the underlying fracture kept propagating after shut-in for the cross-linked gel stage, but not for the slick-water stage.

% - in-line with theoretical predictions of linear hydraulic fracture mechanics \citep{MoLe21}.

We want to derive from the recorded wellhead pressure $p_{wh}$ an approximation of the fracture entry over-pressure $\Delta p_{inj}$, the injection pressure in excess of the in-situ pore pressure. We assume that the in-situ pore pressure is the hydrostatic pressure from the surface, and that, as such, it is balanced out by the hydrostatic pressure in the well. The difference between the well-head and the fracture entry over-pressure then corresponds to the friction losses along the wellbore and at the fracture entry, referred to as $\Delta p_f$. The pressure losses at the fracture entry are usually considered as the sum of two contributions: the perforation friction and the near-wellbore tortuosity friction. We measure at each step the decrease in injection rate $Q_1 \rightarrow Q_2$, the instantaneous decrease of well-head pressure $\Delta p_{wh}$ that we interpret as a decrease in friction losses due to the lesser injection rate. We then fit the following model:

\begin{equation}
    \Delta p_{wh, 1\rightarrow2} = \alpha(Q_1^2 - Q_2^2) + \beta(Q_1^{1/2} - Q_2^{1/2})
\end{equation}

\begin{figure}[ht!]
    \centering
    \includegraphics[width=1\linewidth]{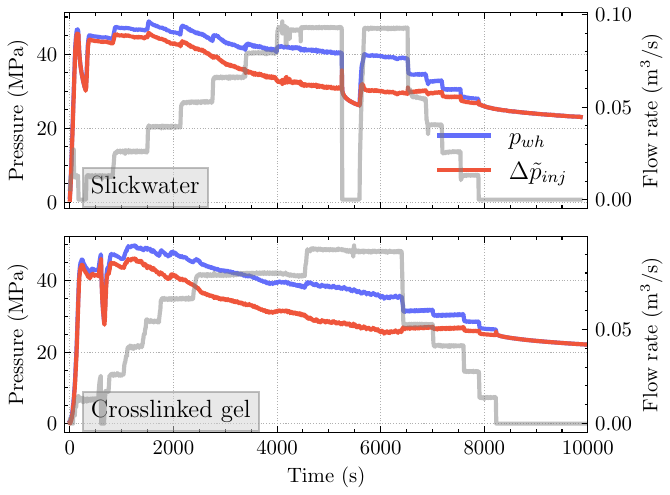}
    \caption{Estimation of pressure friction losses $\Delta p_f$ (wellbore friction and entry friction) as $\Delta p_f = p_{wh} - \Delta p_{inj} = \alpha Q^2 + \beta Q^{1/2}$. Wellhead pressure $p_{wh}$ in blue, modeled fracture entry over-pressure $\Delta \tilde p_{inj}$ in red. The $\alpha$ and $\beta$ parameters are obtained by fitting this model to the immediate pressure drops following flow-rate steps down.}
    \label{fig:friction_losses}
\end{figure}

The first term, proportional to $Q^2$, models the sum of the wellbore and the perforation friction, and the second term, proportional to $Q^{1/2}$, models the near-wellbore tortuosity friction \citep{economides1989}. We obtain similar values for both stages: $\alpha \simeq 9.5$ MPa.m$^{-6}.s^2$ and $\beta \simeq 76$ MPa.m$^{-3/2}.s^{1/2}$ for the slick-water stage; $\alpha \simeq 12$ MPa.m$^{-6}.s^2$ and $\beta \simeq 73$ MPa.m$^{-3/2}.s^{1/2}$ for the cross-linked gel stage. The comparison of observed and modeled immediate pressure drop following rate step down are given in Appendix \ref{Appendix:friction_losses}. We then model the friction losses throughout the injection as $\Delta p_f = \alpha Q^2 + \beta Q^{1/2}$, which we retrieve to the well-head pressure to obtain an approximation of the fracture entry over-pressure $\Delta \tilde p_{inj}=p_{wh}-\Delta p_f$. The resulting modeled fracture entry over-pressure is given in Fig. \ref{fig:friction_losses}. We obtain friction losses of about 10 MPa at the highest injection of 0.1 m$^3$/s for both stimulation stages. 

\subsection{Using the located microseismic events as a proxy of the fracture extent}
\label{section:data:seismic_radius}

\begin{figure}[ht!]
    \centering
    \includegraphics[width=1\linewidth]{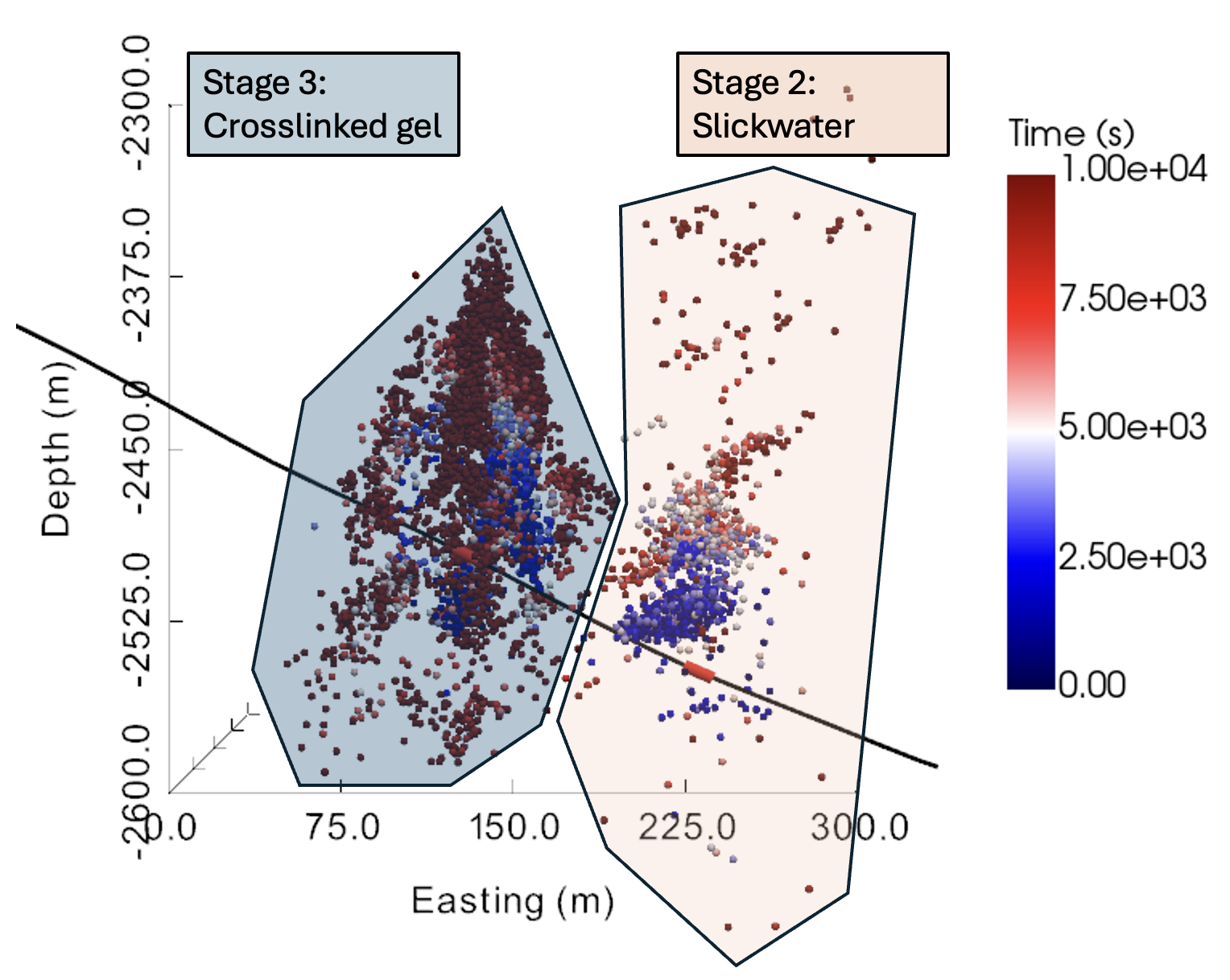}
    \caption{Induced microseismicity catalog associated with the two hydraulic stimulation stages, from \cite{GES2022}. The events are colored by time, with the start of injection of their respective stimulation stage as the origin.}
    \label{fig:3D_seismic_point_cloud}
\end{figure}

In the absence of other geodetic measurements (such as fiber-optics strain distributed sensing or tiltmeters), the induced microseismicity is the only way of investigating how the fractures have developed during the stimulation. The number of recorded and located events for the two stages, as well as the total seismic moments, are reported in Table \ref{tab:seismicity_catalog}. Note that for both stages the moment magnitudes $M_w$ of the localized microseismic events are comprised between -2.0 and 0. An important caveat for direct comparison between the two stages is that the geophones in one of the three vertical wells were out of service during the slick-water stage. 
As such, the catalog of recorded events is likely to be more complete for stage 3, and fewer of the detected events were localized for stage 2 (24\% of the events, accounting for 11\% of the seismic moment) than for stage 3 (40\% of the events, accounting for 44\% of the seismic moment). The uncertainty on the location is also higher for stage 2 (the slick-water stage) than for stage 3 (the median value of the location uncertainty is 34.6 m for stage 3 and 45.4 m for stage 2 as reported in \cite{GES2022}). The spatial and temporal distribution of located events for both stages is shown in Figure \ref{fig:3D_seismic_point_cloud}.

\begin{table*}
\fontsize{10}{12}\selectfont \caption{Induced microseismicity catalog from \cite{GES2022}}
\label{tab:seismicity_catalog}
\centering{}%
\begin{tabular}{p{2cm}p{3cm}p{2cm}p{2cm}p{2cm}p{2cm}p{2cm}}\toprule 
Stage & Number of recorded events & Total seismic moment & Maximum moment magnitude & Number of located events & Total seismic moment of located events & Median value of location uncertainty\tabularnewline
\midrule
slick-water & 5456 & 3.5 10 $^{11}$ N.m & 1.29 & 1322 & 4.0 10$^{10}$ N.m & 45.4 m\tabularnewline
\midrule
cross-linked gel & 13135 & 9.3 10 $^{11}$ N.m & 1.26 & 5283 & 4.1 10$^{11}$ N.m & 34.6 m\tabularnewline
\bottomrule
\end{tabular}
\end{table*}

We assume that most of the located microseismic events are spatially associated with fractures generated during the hydraulic stimulation. These co-located events include not only cracking directly linked to the propagation of fluid-induced tensile or shear fractures, but also the reactivation of pre-existing fractures due to stress changes induced by the advancing main fracture. Given the uncertainty in microseismic event locations, it is not possible to distinguish between these mechanisms.
% We do not restrict this assumption to the fracture tip; microseismicity is expected to occur at any location within the fracture footprint throughout its propagation. 
% We first assume that the created fractures are radial, and will first assess whether the spatial distributions of microseismicity is consistent with this assumption, i.e. that the events are distributed within a disk. We then estimate the evolution in time of the "seismic radius", defined as the sphere that contains the emissions at a given time. If the first assumption is verified, we consider that the seismic radius is a measure of the actual fracture radius.
We propose here a protocol that aims to filter the microseismic events that are co-located with the fluid-induced fractures. Once done, we compute the evolution in time of a "seismic radius": the size of the point cloud formed by these events. Under the assumption that a single planar fracture developed, we finally use this seismic radius as a proxy of the fracture radius.

We start by clustering the located events based on their 3D coordinates using the Density-based Spatial Clustering of Applications with Noise (DBSCAN) algorithm, only keeping the events belonging to the largest cluster. This allows us to have a spatially continuous point cloud that aligns with the hypothesis that these events are a proxy of an underlying quasi-static fracture that is continuous itself. The DBSCAN algorithm uses two parameters: the minimum number of points to form a cluster and the maximum distance between two points to belong to the same cluster, or neighborhood radius. In our use case, focusing only on the more populous clusters, only the neighborhood radius affects the results. 
We choose for its value the median value of the uncertainty on the microseismic events location: 34.6 m for stage 3 and 45.4 m for stage 2.

This leads to eliminate about 1\% of the events during the cross-linked gel stage and 10\% of the events during the slick-water stage. This difference could be a consequence of the difference in acquisition coverage between the two stages.

We then compute a singular value decomposition (SVD) of the selected events 3D coordinates. This decomposition gives us the events mean plane and a quantitative measure of the microseismicity planarity. Noting $\lambda_1 > \lambda_2 > \lambda_3$ the 3 singular values obtain we define the surface variation metric as $\text{SV} = \lambda_3 / (\lambda_1 + \lambda_2 + \lambda_3)$. A surface variation close to zero means a highly planar distribution. By doing so, we quantify that the events that occurred during the cross-linked stage are more distributed on a plane, with a surface variation of about 0.15 compared to the events that occurred during the slick-water stage, which have a surface variation of about 0.3. This difference could also be a consequence of the worse seismic acquisition coverage for the slick-water stage. From these results, we argue that the seismic radius can be used as a proxy of a fracture radius for the cross-linked stage but less so for the slick-water stage.

We note that the events in both stages have the same mean plane orientation; which is sub-vertical (dip of $80^\circ$) and striking N15$^\circ$E. This orientation is very much consistent with the horizontal stress orientation - and therefore with the hypothesis of a HF. However, as shown in Fig \ref{fig:dfn_model_orienttaion}, it is also close to several fracture sets identified by \cite{Finnila2023} - such that reopening / hydro-shearing of natural fractures may not be completely ruled out.

\begin{figure}[ht!]
    \centering
    \includegraphics[width=.7\linewidth]{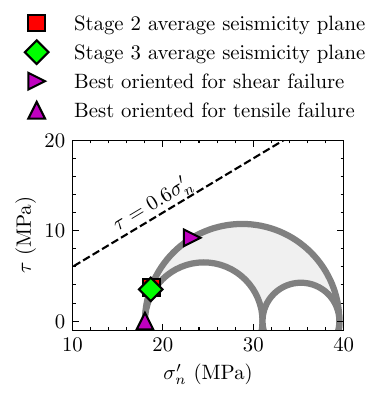}
    \caption{Mohr-circle representation of our "preferred" state of stress model (as reported in Fig. \ref{fig:state_of_stress_model}), along with the projected tractions on the identified mean seismicity planes and on the planes best oriented for tensile and shear failure.}
    \label{fig:mohr_cirlce}
\end{figure}

% Add info on mean fracture planes, very close > hint that both fractures are following the same physics

\begin{figure}[ht!]
    \centering
    \includegraphics[width=1\linewidth]{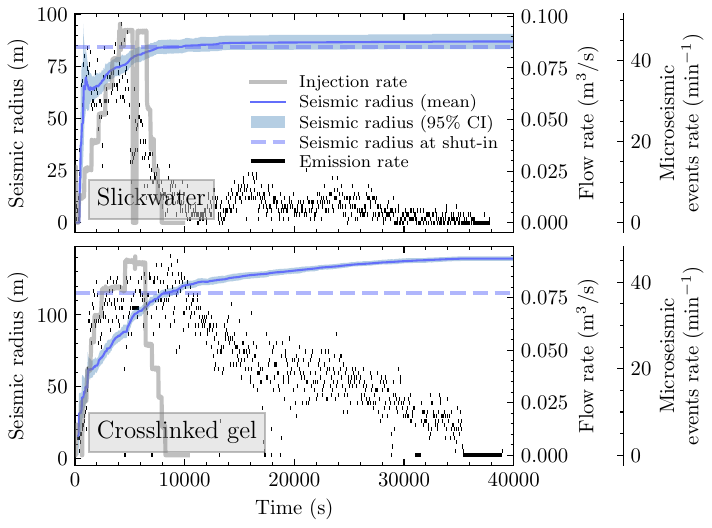}
    \caption{Seismic radius computed from located microseismic events for both stimulation stages. We also show the microseismicity rate (which includes non-located events). In both stages, the microseismicity rate is coherent with the evolution of the seismic radius: for the slick-water stage, the seismic radius stabilizes itself after shut-in, and the microseismicity rate goes down, while for the cross-linked gel, the seismic radius keep on growing after shut-in while the microseismicity rate remains elevated.}
    \label{fig:seismic_radius}
\end{figure}

To compute the seismic radius, we follow the procedure described by \citep{Goebel2018}. At a given time, we consider all the microseismic events that have happened, we compute their barycenter, and we compute the seismic radius as the 95th percentile of the distance to the barycenter of all these events. We follow this procedure for different time intervals to obtain the evolution of the seismic radius with time. We propagate the uncertainty on the events localization through this procedure by using the localization error metric provided in the dataset of \cite{GES2022} and repeating the procedure a large number of times, each time adding to all events a 3D vector with a random unit vector direction and a random norm, drawn uniformly between zero and the provided error. The results are shown in Fig.~\ref{fig:seismic_radius}. We observe that this metric of the fracture evolution with time is coherent with the observations of the evolution of the microseismicity rate after shut-in, we observe post-shut-in growth of the seismic radius for the cross-linked gel stage, but immediate arrest for the slick-water stage.
In Appendix \ref{Appendix:sensitivity_analysis_seismic_radius}, we show a sensitivity analysis of how this resulting seismic radius is affected by the two hyperparameters of our method: the DBSCAN neighborhood radius and the percentile of the distance to the barycenter.

\begin{figure}[ht!]
    \centering
    \includegraphics[width=1\linewidth]{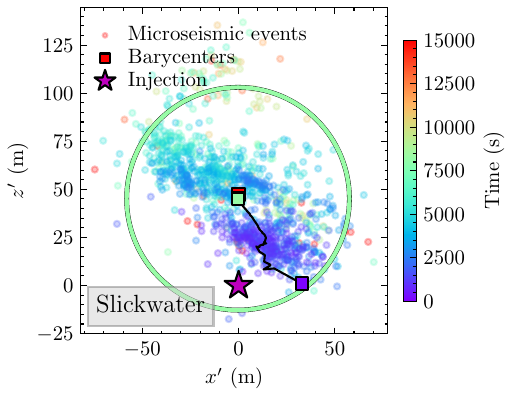}
    \caption{Planar projection of the located seismic emissions on their mean plane, for the slick-water stage. The intersection of this plane with the well, the evolution of the barycenter of the events with time, and the seismic radius at shut-in (green circle) are also represented. $z'$ is taken as the projection of the vertical direction on the mean plane, and the mean plane is itself sub-vertical. Here for the slick-water stage.}
    \label{fig:2d_events_stage2}
\end{figure}

\begin{figure}[ht!]
    \centering
    \includegraphics[width=1\linewidth]{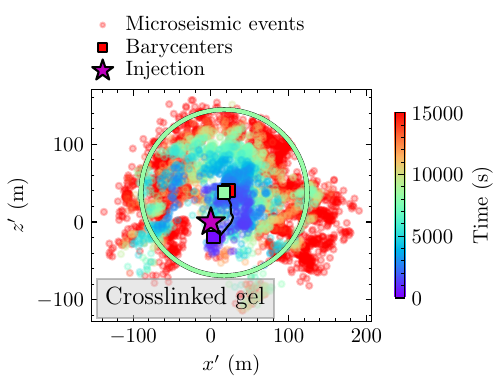}
    \caption{Planar projection of the located seismic emissions on their mean plane, for the cross-linked gel stage.  The intersection of this plane with the well, the evolution of the barycenter of the events with time, and the seismic radius at shut-in (green circle) are also represented. $z'$ is taken as the projection of the vertical direction on the mean plane, and the mean plane is itself sub-vertical. Here for the cross-linked gel stage.}
    \label{fig:2d_events_stage3}
\end{figure}

We are confident in assimilating this seismic radius with a fracture radius for the cross-linked gel stage due to its higher planarity (seen visually and quantified via the surface variation metrics) but less so for the slick-water stage. In Figures \ref{fig:2d_events_stage2} and \ref{fig:2d_events_stage3}, we show for both stages the projected events on their mean plane (the projection maintains the vertical), along with the evolution of the computed barycenter with time and the seismic radius at shut-in. This reveal the a rather radial distribution of the microseismicity for the cross-linked stage. This is less the case for the slick-water stage.

% \begin{table*}
% % \fontsize{10}{12}\selectfont \caption{Induced microseismicity catalog from \cite{GES2022}}
% \label{tab:seismicity_catalog}
% \centering{}%
% \begin{tabular}{p{2cm}p{3cm}p{3.5cm}}\toprule 
% Stage & Surface variation & Visually distributed on a plane \tabularnewline
% \midrule
% slick-water & 0.27 & No \tabularnewline
% \midrule
% cross-linked gel & 0.15 & Yes \tabularnewline
% \bottomrule
% \end{tabular}
% \end{table*}

% Figure \ref{fig:3D_seismic_point_cloud} shows the point clouds formed by the microseismic events that have been successfully located for the two stimulation stages.

% \begin{figure}[ht!]
%     \centering
%     % \includegraphics[width=1\linewidth]{Figures/Misc/seismic_point_cloud_visu_assembled.png}
%     \includegraphics[width=1\linewidth]{Figures/Misc/GES2022_3D_point_cloud_annotated.png}
%     \caption{Induced microseismicity catalog associated with the hydraulic stimulation, from \cite{GES2022}. The events are colored by time, with the start of injection of their respective stimulation stage as the origin.}
%     \label{fig:3D_seismic_point_cloud}
% \end{figure}
% ----------------------------------
% METHODS
% ----------------------------------
\section{Analytical and numerical modeling}

Our modeling approach first relies on well-established analytical models, for which scaling laws and asymptotic solutions for the fracture extent are available, describing both planar radial hydraulic fractures and planar fluid-induced frictional ruptures.

We then validate the insights derived from these analytical models using numerical simulations performed with a fully coupled hydro-mechanical solver. 

In this section, we first describe the analytical models for tensile hydraulic fractures and fluid-induced shear fractures. We then provide a brief description of   our fully-coupled numerical model.

\subsection{Analytical model for growth and arrest of radial hydraulic fractures}
\label{section:methods:hf_model}

A review of closed-form developments regarding the problem of a radial hydraulic fracture can be found in \cite{Detournay2016}. In this planar penny-shaped HF model, the following assumptions are made:
i) the medium is an infinite isotropic linear elastic solid, ii) the fracturing fluid is Newtonian and injected from a point source at a constant rate, and iii) the flow within the fracture follows the lubrication approximation. Quasi-static linear elastic fracture mechanics (LEFM) theory is assumed. The host rock is either considered impermeable or permeable under Carter's leak-off approximation: one-dimensional and pressure-independent leak-off. The main result is that the fracture propagation is governed by competition between viscous flow dissipation and energy expanded in the creation of new fracture surfaces on one hand and competition between fluid storage in the fracture and leak-off in the surrounding rock on the other hand. Four limiting propagation regimes follow:
\begin{itemize}
    \item The viscosity-storage dominated regime, referred to as the $\mathcal{M}$-vertex
    \item The toughness-storage dominated regime, referred to as the $\mathcal{K}$-vertex
    \item The viscosity-leak-off dominated regime, referred to as the $\tilde{\mathcal{M}}$-vertex
    \item The toughness-leak-off dominated regime, referred to as the $\tilde{\mathcal{K}}$-vertex
\end{itemize}

For each of these regimes, scaling laws for the fracture radius as a function of the problem's parameters have been derived by \citep{Detournay2016}. They depend on the injection rate $Q_0$, and the material parameters:
\begin{align}
    E' = \frac{E}{1-\nu^2}, \quad K'=4\sqrt{\frac{2}{\pi}}K_{Ic}, \\ \mu'=12\mu, \quad C'=2C_L
\end{align}
where $E$ is the Young modulus, $\nu$ is the Poisson's ratio, $K_{Ic}$ is the mode I fracture toughness, $\mu$ is the fluid dynamic viscosity, and $C_L$ is the Carter's leak-off coefficient. Among the obtained theoretical results is the time-dependent length scale of the fracture under each of the four propagation regimes.  Under the $\mathcal{M}$-vertex viscosity/storage dominated regime this fracture length scale is:
\begin{equation}
    L_m = \left(\frac{E'V(t)^3t}{\mu'}\right)^{1/9}
    \label{eq:scale_m_vertex}
\end{equation}

% The scalings for the fracture length/radius are given in Table \ref{tab:hf_length_scale}, where we have .

% \begin{table*}
% \fontsize{10}{12}\selectfont \caption{Fracture length scales for the radial HF, from \cite{Detournay2016}}
% \label{tab:hf_length_scale}
% \centering{}%
% \begin{tabular}{p{2cm}p{2cm}p{2cm}p{2cm}p{2cm}}\toprule 
%  Scaling & $\mathcal{M}$ & $\mathcal{K}$ & $\tilde{\mathcal{M}}$ & $\tilde{\mathcal{K}}$ \tabularnewline
% \midrule
% Fracture length scale & $\left(\frac{E'V^3t}{\mu'}\right)^{1/9}$ & $\left(\frac{E'V}{K'}\right)^{2/5}$ & $\left(\frac{V^2}{tC'^2}\right)^{1/4}$ & $\left(\frac{V}{tC'^2}\right)^{1/4}$ \tabularnewline
% \bottomrule
% \end{tabular}
% \end{table*}

Where the injection rate $Q_0$ is replaced by $V(t)/t$, with $V(t)$ the injected volume, as done by \citep{MoLe21}. An important result for our use-case is that under the usual material parameters found for field hydraulic fractures it is expected that the fracture first follows an early-time asymptote (denoted as the $\mathcal{O}$-solution), then transition to the $\mathcal{M}$-solution associated to the viscosity-storage dominated regime, and finally transitions to the $\mathcal{K}$-solution associated to the toughness-storage regime or to the $ \tilde{\mathcal{M}}$-solution associated to the viscosity-leak-off regime. This defines 3 transition times: $t_{om}$, $t_{mk}$ and $t_{m\tilde m}$. Their expressions are given as:
\begin{align}
    t_{om} = \frac{E'^2\mu'}{\sigma'_0}, \quad t_{mk} = \left(\frac{E'^{13}\mu'^{5}Q_0^3}{K'^{18}}\right)^{2}, \notag \\
    t_{m\tilde m} = \left(\frac{\mu'^{4}Q_0^6}{E'^4 C'^{18}}\right)^{1/7}
    \label{eq:transition_times_hf}
\end{align}

The behavior of the fracture after shut-in is relevant in this study. The problem of the growth and arrest of a radial HF upon shut-in has been studied by \cite{MoLe21}. The authors notably show that only fractures propagating in the viscosity-storage dominated regime exhibit post-shut-in growth. In contrast, the fractures propagating in any of the three other limiting regimes arrest immediately at shut-in (with closure of the fracture if considering leak-off). They show that for the $\mathcal M$-vertex regime, the fracture length scale defined in Eq. \ref{eq:scale_m_vertex} holds after shut-in until the fracture arrest. This arrest radius is either governed by the fracture energy criterion or by the fluid-leak-off. Denoting $R_{k,a}$ the arrest radius in the first case and $R_{\tilde m,a}$ the arrest radius in the second case, they obtain the following expressions:

\begin{align}
    R_{k,a} &= \left(\frac{3}{\pi\sqrt{2}}\frac{E'V}{K'}\right)^{2/5}, \notag\\
    R_{\tilde m,a} &= \gamma_{\tilde m,a} \left(\frac{E'V^5}{C'^2 \mu'}\right)^{1/13}
    \label{eq:hf_arrest_radius}
\end{align}
Where $\gamma_{\tilde m,a}$ is a scalar function of the problem's dimensionless leak-off coefficient. \cite{MoLe21} show that for $\mathcal{C}_s < 0.25$, $\gamma_{\tilde m,a}\simeq 0.5218$ with $\mathcal{C}_s$ the dimensionless leak-off at shut-in ($t_s$ the shut-in time) defined as:
\begin{equation}
    \mathcal{C}_s = C'\frac{E'^{2/9}t_s^{7/18}}{Q_0^{1/3}\mu'^{2/9}}
    \label{eq:non_dim_leakoff}
\end{equation}

\subsection{Analytical model for fluid-induced shear fractures}
\label{section:methods:shear_f_model}

We rely on the work of \cite{Saez2022} on planar circular fluid-induced frictional ruptures. Their model supposes a fault $\Gamma$ on the $(x,y)$ plane that separates two semi-infinite isotropic homogeneous linear elastic solids  subjected to a uniform state of stress consisting of a normal component $\sigma_0$ and a shear component $\tau_0$. The fault plane obeys a Coulomb friction yielding law with a constant friction coefficient $f$ and no cohesion. A fluid point source lying on the plane models the injection of fluid from a wellbore crossing the fault plane. Finally, it is supposed that the rock is impermeable and that the fluid flows only within the fault interface with a constant permeability $k$, independent of the mechanical deformation of the system. The hydro-mechanical problem is, as such, said to be one-way coupled. The width-averaged fluid pore-pressure diffusion within the fault follows:
\begin{equation}
    Sw_h\frac{\partial p}{\partial t} - w_h\frac{k}{\mu}\nabla^2p = 0
    \label{eq:saez_pore_pressure}
\end{equation}
Where $S$ is a storage coefficient (combining the compressibility of the fluid and of the porous material within the fracture width, $\mu$ is the fluid viscosity, and $w_h$ is the fault hydraulic width (considered constant as well). This defines a diffusion coefficient $\alpha = k/\mu S$. This equation in an axisymmetric setting for an infinite medium with a constant rate point source injection has an analytical solution as given in \citep{Theis1935}. This solution exhibits a characteristic pressure value: the overpressure at the injection point $\Delta p^*$:
\begin{equation}
    \Delta p^* = \frac{Q_0 \mu}{4\pi kw_h}
    \label{eq:theis_dpstar}
\end{equation}

Where $Q_0$ is the injection rate. Regarding the mechanical part of the problem, the authors show that the solution to this problem is self-similar and that the rupture radius $R(t)$ (or slip front position) follows:
$$
R(t) = \lambda  \sqrt{4\alpha t}
$$
where $\sqrt{4\alpha t}$ is a characteristic distance for the pore-pressure diffusion and $\lambda$ is a function of a unique dimensionless coefficient $T$ referred to as the fault stress injection parameter : 
$$
T = \frac{f\sigma_0 - \tau_0}{f\Delta p*}
$$

The authors identify two limiting regimes based on the value of $T$: a marginally pressurized regime corresponding to $T \approx 1$ (for which $\lambda \ll 1$, and a critically stressed regime corresponding to $T\ll1$ (for which $\lambda \gg 1$). In \cite{Saez2023}, the authors look at the post-shut-in behavior of such fluid-induced shear rupture. They show that post-shut-in growth is expected under the critically stressed regime, while immediate arrest upon shut-in is expected under the marginally pressurized regime.

\subsection{Numerical solver of a fully coupled hydro-mechanical mix-mode fluid-induced fracture model}
\label{section:methods:pyfracx_model}

To model the hydraulic stimulation stages without any assumption about their nature (being tensile hydraulic fractures or fluid-induced dilatant shear ruptures), we solve the full coupled hydro-mechanical problem, accounting for shear and tensile failure of fractures. The model has two variables defined over the fracture domain: the pore-pressure $p(\bm{x},t)$ and the displacement discontinuity field $\bm{u}(\bm{x},t) = \bm{u}^+ - \bm{u}^-$ with $\bm{x} \in \Gamma$, $\Gamma$ being a set of preexisting interfaces (or fractures) within the rock. These two fields $p$ and $\bm{u}$ verify a system of 3 equations: the elastic equilibrium, the fluid diffusion, and the interface constitutive equation relating the tractions to the displacements on the fracture. Assuming the host rock to be an infinite linear isotropic elastic solid, we rely on the boundary element method (BEM) to solve for the elastic equilibrium. We consider that the width-averaged fluid flow follows the lubrication approximation and comes in two flavors given in Eq. \ref{eq:pure_cubic_law} and \ref{eq:shear_zone_cubic_law} referred to as the cubic law and the shear-zone cubic law, respectively. Here $u_n$ is the mechanical opening (due to tensile opening or shear-induced dilation), $S$ is a the uniaxial content strain storage coefficient (taken as the compressibility of the fluid in the pure cubic law), $u_{n,0}$ and $T_0$ are initial hydraulic opening and initial transmissibility respectively. The main difference is that the shear-zone cubic law allows us to set the storage to be about constant by having the $h$ parameter, akin to a fault zone width, very large compared to the expected mechanical opening of the fracture $u_n$. In what follows, we will use Eq. \ref{eq:pure_cubic_law} when modeling a pure hydraulic fracture problem (where we suppose that a new tensile fracture opens) and Eq. \ref{eq:shear_zone_cubic_law} when modeling existing fracture reactivation:
\begin{align}
    S u_n \frac{\partial p}{\partial t} - \nabla.\left(\frac{(u_{n,0} + u_n)^3}{12\mu}\nabla p\right) = \gamma 
    \label{eq:pure_cubic_law}\\
    S (h + u_n) \frac{\partial p}{\partial t} - \nabla.\left(\frac{T_0 + u_n^3/12}{\mu}\nabla p\right) = \gamma
    \label{eq:shear_zone_cubic_law}
\end{align}

The third element of this model is the interface constitutive relation, which relates the effective traction $\bm T' = \bm T + p\bm n$ ($\bm T$ being the total traction, $p$ the pore-pressure, and $\bm n$ a normal unit vector to the fracture surface) to the displacement discontinuity $\bm u$. We use an elasto-plastic constitutive relation that describes fractures that would first deform elastically, given shear and normal interface stiffness parameters. When the failure criterion is reached, the interface starts to deform in an inelastic (referred to as plastic) manner. In this regard, the displacement discontinuity is decomposed into an elastic and a plastic part: $\bm u = \bm u^e + \bm u^p$. Two failure criteria are introduced, one for shear failure $F_{\text{shear}}$ and one for tensile failure $F_{\text{tensile}}$, each of them is associated with a plastic flow rule to compute the inelastic displacements when the corresponding failure criterion is reached in a way to ensure that the tractions remain on the yielding surface. The parameters that define these failure criteria and the plastic flow rules depend on internal variables to account for the damage of the interface: a shear damage variable $\chi_s$ and a total damage variable $\chi_t$ are defined respectively as the maximum encountered of the shear part and the norm of the encountered inelastic displacements. The shear failure criterion is the Mohr-Coulomb criterion with possibly changing friction coefficient and cohesion as defined in \ref{eq:pyfracx_shear_fc} where $\bm\tau$ is the shear component of $\bm T'$ and $T'_n$ is its normal component. In our model, the friction coefficient drops from an initial value $f_p$ to a residual value $f_r < f_p$ over a slip weakening distance $\delta_c$. The cohesion $C$ is taken as decreasing from a peak value to zero over a critical tensile weakening distance $w_c$. The tensile failure criterion is defined as \ref{eq:pyfracx_tens_fc} where tensile failure is reached when the normal effective traction reaches the tensile strength of the material $\sigma_t$, $\sigma_t$ is taken as a degrading from an initial value $\sigma_{t,p}$ to zero over a critical tensile weakening distance $w_c$. Note that $T'_n$ is taken positive in tension.
\begin{align}
    F_{\text{shear}}(\bm T', \bm\chi) = ||\bm\tau|| - f(\chi_s) T'_n - C(\chi_t)
    \label{eq:pyfracx_shear_fc} \\
    F_{\text{tensile}}(\bm T', \bm\chi) = T'_n + \sigma_t(\chi_t)
    \label{eq:pyfracx_tens_fc}
\end{align}
Note that the degradation of the friction coefficient and the tensile strength with the inelastic displacement discontinuity define a shear and a tensile fracture energy. In the tensile case, this fracture energy is $\sigma_{t,p}w_c/2$ and hence the associated fracture toughness is $K_{Ic} = (E \sigma_{t,p}w_c/2)^{1/2}$. This implementation is akin to a cohesive zone model. When the failure criteria are reached. For the tensile failure case, we use an associated plastic flow rule, whereas for the shear failure case, we use a non-associated plastic flow rule to account for dilatancy. This dilatancy is modeled as: 
\begin{equation}
    \dot u_n = \tan(\psi)(\chi_s) \dot{||\bm\delta||}
    \label{eq:pyfracx_dilatancy}
\end{equation}
Where $\dot u_n$ is the normal opening velocity, $\bm\delta$ is the slip velocity, and $\tan(\psi)$ is a dilatancy coefficient chosen to drop from an initial value $\tan(\psi_p)$, $\psi_p$ being the peak dilation angle, to zero over the slip weakening distance $\delta_c$. The mechanical opening $u_n$ that controls the fluid flow can then either be caused by pure tensile opening of by shear-induced dilation. 

Finally, we can choose to relax the hypothesis that the host rock is impermeable and introduce fluid leak-off. This is implemented as one-dimensional pressure-dependent leak-off. The pore pressure in the bulk domain follows a constant storage/constant permeability Darcy flow diffusion equation: 
\begin{equation}
    S_b \frac{\partial p}{\partial t} - \frac{k_b}{\mu_b}\nabla^2 p = \gamma
    \label{eq:bulk_flow}
\end{equation}
where the $b$ subscript is used to refer to properties of the bulk domain. Carter's leak-off model was not suited for our case as the pressure-independent approximation would not hold for the pressure profiles expected in fluid-induced dilatant shear fracture problems. The relation between one-dimensional pressure-dependent leak-off and Carter's leak-off model is discussed in \cite{Kanin2020}.

Regarding the numerical implementation, we discretize the fracture domain geometry and then rely on a boundary element method formulation for the rock elastic equilibrium problem and a finite element method formulation for the fluid diffusion problem (both in the fracture domain and the bulk domain if considering leak-off). The interface constitutive equation is local, it is integrated at every node of the mesh at each solver's iteration. The problem is highly non-linear, due to the cubic dependence of the fracture hydraulic transmissibility on the fracture opening. We rely on a fully implicit monolithic solver, base on a Newton-Raphson iteration at each time-step to solve the coupled hydro-mechanical systems. 

The model entries are : 
\begin{itemize}
    \item The initial in-situ stresses before fluid injection $\sigma_{ij}^o$
    \item The elastic moduli of the rock: the Young's elastic modulus $E$ and the Poisson's ratio $\nu$
    \item The initial hydraulic properties of the interface: its storage and conductivity (defined slightly differently depending on the constitutive law used: pure cubic law or shear zone cubic law)
    \item The fluid dynamic viscosity $\mu$
    \item The interface constitutive law parameters: its stiffness components $k_s$ and $k_n$ (for shear and normal elastic deformation), the shear failure parameters ($f_p$, $f_r$, $C$, $\psi_p$, and $\delta_c$), and the tensile failure parameters ($\sigma_t$ and $w_c$)
    \item If considering leak-off, the storage and conductivity $\kappa = k_b/\mu_b$ of the bulk domain
    \item The fractures geometry
\end{itemize}

This solver has been validated against known analytical solutions for both hydraulic fracture problems and fluid-induced shear fracture problems. Additional information regarding the solver and its validation can be found in \cite{LeGu25}.

% ----------------------------------
% RESULTS
% ----------------------------------
\section{Modeling of FORGE 2022 hydraulic stimulations}

We apply the modeling framework introduced above to the 2022 Utah FORGE hydraulic stimulation experiments. We first analyze the stimulations under the assumption of tensile hydraulic fracturing, and then investigate the alternative hypothesis of fluid-induced dilatant shear fracturing. In both cases, simulations are conducted using the preferred in situ stress state introduced in Section \ref{section:data:state_stress}. Finally, we propagate uncertainties in the stress state and in natural fracture orientations through the numerical model to assess their impact on the results.

\subsection{Modeling the stimulations as hydraulic fractures}
\label{section:results:as_hf}

We start here by going back into the hydraulic stimulation data under the hypothesis that these are planar radial hydraulic fractures , hence making use of the results introduced in Section \ref{section:methods:hf_model} as well as the solver introduced in Section \ref{section:methods:pyfracx_model}. This hypothesis is highly supported by the orientation of the mean plane of the seismicity (almost identical for both stages) and especially for the cross-linked stage, where the geometry of the well-resolved induced microseismicity allows us to confidently say that the rupture was indeed planar and radial (to the first order).

 Making use of the results of \cite{MoLe21} introduced before, we can now interpret the difference in the post-shut-in evolution of the microseismicity rate between the two stages: that the rate immediately decreases following shut-in for the slick-water stage, while it remains elevated for a few hours for the cross-linked gel stage. We recall that from our seismic radius analysis in Section \ref{section:data:seismic_radius}, this difference in induced seismicity activity translates to a difference in post-shut-in evolution of the seismic radius that immediately arrests in the slick-water case but keeps propagating in the cross-linked gel. Under the HF hypothesis, this could be explained by the fracture propagating in the $\mathcal{M}$-vertex viscosity-storage dominated regime in the cross-linked gel case, which exhibits post-shut-in growth, but in a regime that does not exhibit such post-shut-in growth in the slick-water case, due either to an elevated toughness or to an elevated leak-off or both. This is coherent with how the transition times $t_{mk}$ and $t_{m\tilde m}$ scale with the viscosity $\mu$, as recalled in Eq. \ref{eq:transition_times_hf}. Given that the cross-linked gel has a viscosity roughly 3 orders of magnitude greater than the slick-water, it is expected that the ratio of $t_{mk}$ times to be of the order of $10^8$ while the ratio of $t_{m\tilde m}$ times to be of the order $10^2$. Said differently, the higher viscosity of the cross-linked gel would allow the hydraulic fracture to remain in the viscosity-storage dominated regime within the treatment time, while, when using slick-water, the fracture would have time to transition to a toughness or leak-off dominated regime.

\begin{figure*}[ht!]
    \centering
    \includegraphics[width=.7\linewidth]{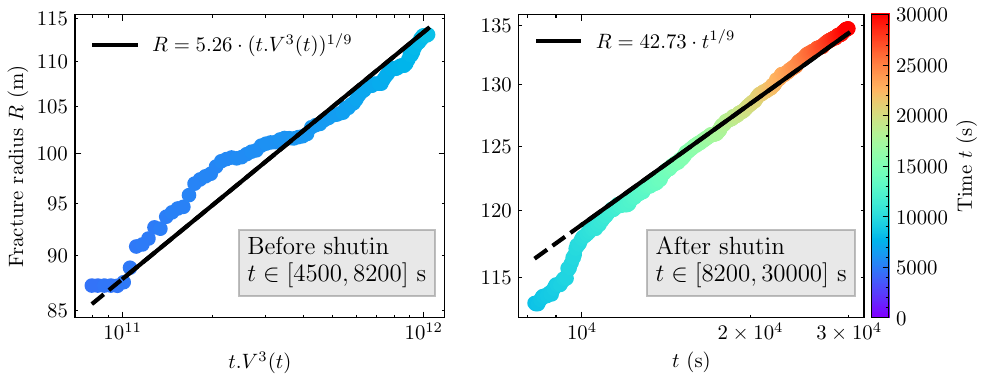}
    \caption{Recovering the fracture length scaling expected for a viscosity-storage dominated hydraulic fracture, recalled in Eq. \ref{eq:scale_m_vertex}, in the fracture radius computed from the located microseismic events, for the cross-linked gel stage. The left figure shows that we recover the expected scaling during injection for the second half of the stimulation, and the right figure shows that we recover the expected scaling in the post-shut-in growth. We do not recover the expected scaling during the first half of the stimulation ($t < 4500$ s), see Appendix \ref{Appendix:stage3_Mvertex_scaling} for the associated data.}
    \label{fig:recovering_mvertex_scaling}
\end{figure*}

For the cross-linked gel stage, we can push the analysis further using the fact that we can confidently use the computed seismic radius in Section \ref{section:data:seismic_radius} as a proxy of the fracture radius, and compare this fracture radius to the expected scaling of such $\mathcal{M}$-vertex propagation regime. As recalled in Eq. \ref{eq:scale_m_vertex}, we expect the fracture radius to scale linearly with $(V(t)^3t)^{1/9}$ ($t$ the time and $V(t)$ the injected volume), both before and after shut-in. We show that we recover this scaling both before and after shut-in, as shown in Figure \ref{fig:recovering_mvertex_scaling}. Before shut-in we only recover the scaling starting from the start of the highest injection rate step around $4500$ s. Observing this match between the seismicity-inferred fracture radius and the expected theoretical scaling is strong evidence in favor of the planar radial HF hypothesis for the cross-linked gel stimulation stage. We do not interpret the fact that we don't recover the expected scaling before $4500$ s, the associated plot is given in Appendix \ref{Appendix:stage3_Mvertex_scaling}. 
These results are robust regarding the choice of the two hyperparameters used to compute the fracture radius from the located microseismic events, as shown in Fig. \ref{fig:recovering_mvertex_scaling_ensitivity_analysis}.

We then propose to run numerical simulations for both fracturing fluids (and injection rate schedules), first assuming that the rock is impermeable. We use the numerical solver introduced in \ref{section:methods:pyfracx_model} using the parameters reported in Table \ref{tab:material_properties}. All numerical simulations are performed in a 3D axisymmetric framework, consistent with the assumption of planar radial fractures. This axisymmetric formulation has two main limitations. First, it precludes the inclusion of stress gradients; as a result, the upward bias in fracture propagation observed in the microseismicity cannot be reproduced. Second, it requires the Poisson’s ratio to be set to zero, which has no effect on tensile hydraulic fractures modeling and is expected to have only a minor influence on shear fracture geometry and not constitute a first-order effect \citep{Saez2022}.  An important caveat of the numerical modeling of the cross-linked gel stage is that we consider the fluid in our simulations to be Newtonian with a constant viscosity, while the cross-linked gel is expected to have a shear-thinning non-Newtonian rheology and, more importantly, its effective viscosity is expected to decrease over the duration of the stimulation due to thermal degradation of its constitutive polymer. 
We investigate the effect of the chosen value of viscosity on our results through a sensitivity analysis (Section \ref{section:incluence_viscosity}). We show that our results are robust to this choice of viscosity value. We have confidence that the use of a constant viscosity Newtonian rheology, with results robust to the choice of the viscosity value, allow us to effectively model the stimulations to the first-order. A theoretical treatment of the propagation of an hydraulic fracture by a fluid exhibiting such temperature-dependent, possibly non-Newtonian, rheology would be required to further validate this claim.

\begin{table}
\begin{threeparttable}
\fontsize{10}{12}\selectfont \caption{Material properties used for numerical simulations assuming a pure tensile hydraulic fracture within an impermeable medium}
\label{tab:material_properties}
\centering{}%
\begin{tabular}{ll}
\toprule 
Young's modulus $E$ & 42 GPa\tabularnewline
Poisson's ratio $\nu$\tnote{(1)} & 0.0\tabularnewline
Mode I toughness $K_{Ic}$\tnote{(2)} & 3.6 MPa.m$^{1/2}$\tabularnewline
Eff. normal stress $\sigma_0$\tnote{(3)} & 18.0 MPa\tabularnewline
Water compressibility (both fluids)\tnote{(4)} & $6.10^{-10}$ Pa$^{-1}$ \tabularnewline
Viscosity $\mu$ (slick-water)\tnote{(4)} & 1.3.10$^{-4}$ Pa.s\tabularnewline
Viscosity $\mu$ (cross-linked gel)\tnote{(5)} & 0.1 Pa.s\tabularnewline
\bottomrule
\end{tabular}
\begin{tablenotes}
  \small
  \item[1] Imposed by the axisymmetric setting
  \item[2] Twice the lab value
  \item[3] From our "preferred" state of stress
  \item[4] From the IAWPS at reservoir conditions ($220^\circ$ and $25$ MPa)
  \item[5] Aligned with other studies ($0.07$ Pa.s in \cite{McClure2023}, $[0.1,0.2]$ Pa.s in \cite{Xing2021}). 
  
\end{tablenotes}
\end{threeparttable}
\end{table}

\begin{figure}[ht!]
    \centering
    \includegraphics[width=1\linewidth]{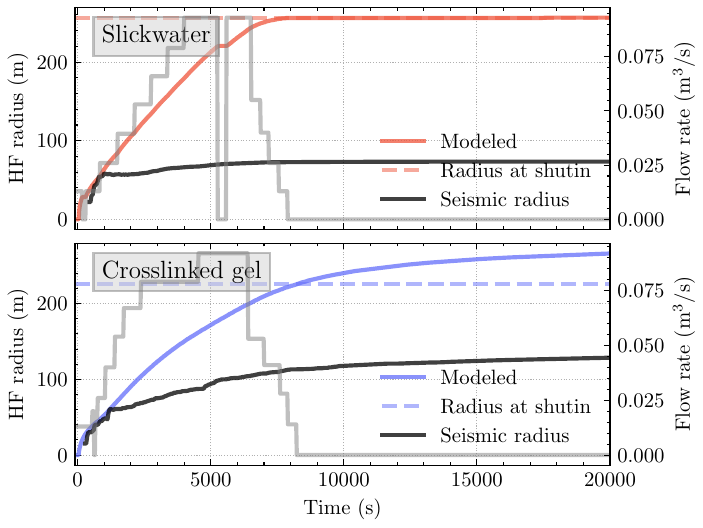}
    \caption{Modeled HF radius for the two stimulation stages, differing by the fracturing fluid viscosity and the flow rate schedule. The parameters used are reported in Table \ref{tab:material_properties}. No fluid leak-off is considered here.}
    \label{fig:newHF_noleakoff}
\end{figure}

\begin{figure*}[ht!]
    \centering
    \includegraphics[width=.7\linewidth]{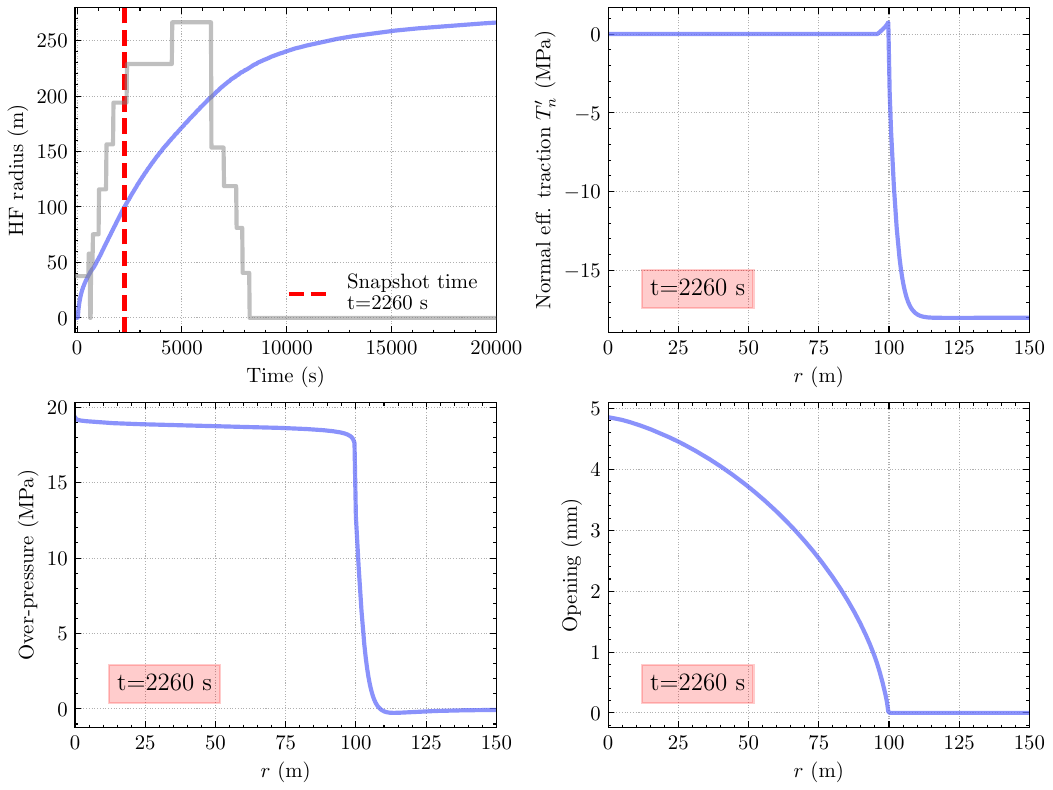}
    \caption{Detailed view of the numerical solver output: evolution of the fracture radius with time, along with the profile of opening, over-pressure, and normal effective traction at $t=2260$ s. $r$ is the radial coordinate with $r=0$ the injection point. The model includes diffusion ahead of the fracture front, but the associated fluid loss is minimal; we still recover the known analytical solutions for radial HF \citep{LeGu25}.}
    \label{fig:newHF_snapshot}
\end{figure*}

Figure \ref{fig:newHF_snapshot} shows a snapshot view at a given time of the solution for the cross-linked gel stage. It shows the problem two variables: the opening, or the normal component of the displacement discontinuity vector, and pressure, as well as the normal effective traction. This is shown to better illustrate what the model solves for. In what follows, we will only show the modeled fracture radii and injection pressures, which are the two quantities we have observations of. The evolution of the resulting HF radius for both fracturing fluids is shown in Figure \ref{fig:newHF_noleakoff}. We see that for these values of viscosity and fracture toughness, we recover the immediate arrest after shut-in in the slick-water case while obtaining post-shut-in growth in the cross-linked gel case. We note that under these parameters and the impermeable rock hypothesis, the predicted fracture radii are greatly overestimated compare to our estimated values from the located microseismicity.

We rely on the work of \citep{MoLe21}, introduced in section \ref{section:methods:hf_model}, to evaluate the toughness or the leak-off required to match the observed final fracture radius for the cross-linked gel stage, that is of about $110$ m, using the expressions introduced in Eq. \ref{eq:hf_arrest_radius}. The other parameters controlling the fracture arrest radius (elastic modulus and viscosity) are better constrained and do not have a strong effect on the fracture radius within their range of expected values. We obtain the following result: if we suppose that the fracture arrested solely due to fracture energy we would need a toughness of $K_{Ic} \simeq 22$ MPa.m$^{1/2}$, while if we suppose that the fracture arrested solely due to leak-off we would need a Carter's leak-off coefficient of $C_L \simeq 10^{-4}$ m/s$^{1/2}$. Both these values appear unrealistically elevated; this is clear evidence that neither the zero-leak-off nor the zero-toughness asymptotic solutions, used in Eq. \ref{eq:hf_arrest_radius}, allows to reproduce the post-shut-in growth of the fracture in the cross-linked gel stage. It indicates that leak-off can not be neglected entirely, and significantly reduced the fracture size.

\begin{figure}[ht!]
    \centering
    \includegraphics[width=1\linewidth]{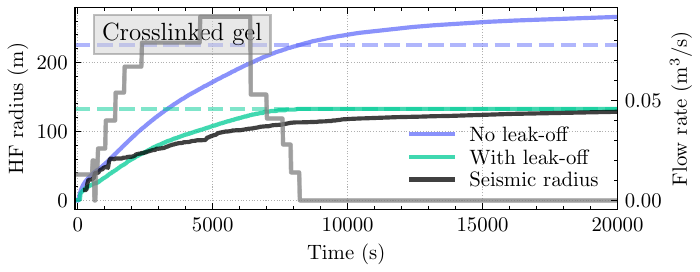}
    \caption{Accounting for fluid leak-off during the cross-linked stage allows to match the observed fracture radius but do not allow to reproduce the post-shut-in growth.}
    \label{fig:newHF_stage3_addingLeakOff}
\end{figure}

\begin{figure}[ht!]
    \centering
    \includegraphics[width=1\linewidth]{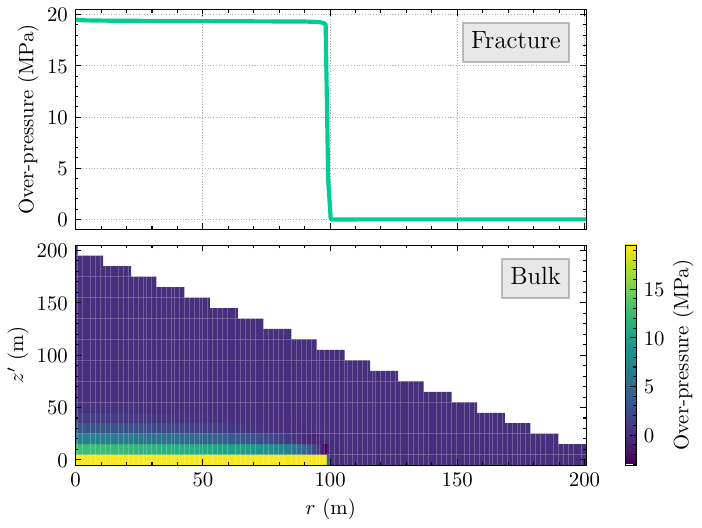}
    \caption{Snapshot view of the over-pressure within the fracture and the bulk domains at $t=2260$ s.} % Here it should be said that the diffusion length of the pore pressure is chosen so to be resolved by the mesh.
    \label{fig:newHF_stage3_leakoff_2dplot}
\end{figure}

Figure \ref{fig:newHF_stage3_addingLeakOff} displays the HF radius obtained when adding leak-off to the cross-linked gel stage simulation. We keep the same value of toughness, of $K_{Ic} = 2K_{Ic,lab} \simeq 3.6$ MPa.m$^{1/2}$, and we adjust the leak-off so to match the observed fracture radius. In Figure \ref{fig:newHF_stage3_leakoff_2dplot}, we show a snapshot view of the associated pore pressure both in the fracture and in the bulk (surrounding rock). In our model, the leak-off is parametrized by the bulk conductivity $\kappa_b = k_b/\mu_b$  and storage $S_b$. The conductivity encompasses both the permeability of the rock and the viscosity of the fluid leaking-off, this viscosity $\mu_b$ combines the fracturing fluid viscosity (here the cross-linked gel) and the viscosity of the water that originally fills the bulk. If we take $\mu_b$ as the viscosity of the water, its lower bound, and $S_b$ as the compressibility of water times a porosity of 1\%, we obtain an estimated permeability of the surrounding rock mass to be about $k_b\simeq 3.10^{-16}$ m$^2$. This value is an upscaled permeability, given that the intact rock matrix can be considered impermeable on these timescales, it is to be related to the permeability of the natural fracture networks. This value is aligned with the upper bound of the upscaled permeability of \cite{Finnila2023b}, computed given a DFN model of the Utah FORGE reservoir and assumptions on the hydraulic conductivity of these fractures. The assumption that the fluid leak-off effective viscosity is close to the water viscosity is not so irrelevant given that the vast majority of the fluid in the rock is water and we only have cross-linked gel close to the fracture. We can also compute an equivalent Carter's leak-off coefficient as $C_L= \sigma_0'\sqrt{\frac{S_b k_b}{\pi \mu_b}}$ here evaluated to $C_L \simeq 10^{-5}$ m/s$^{1/2}$. If adding leak-off allows us to match the final fracture radius, we observe that this amount of leak-off and the homogeneous leak-off assumption lead to the immediate arrest of the fracture after shut-in, and do not reproduce the observed post-shut-in growth. 

A possible explanation to explain the limited fracture extent due to leak-off and the post-shut-in growth is polymer filter cake buildup: the polymer in the cross-linked gel may accumulate within the rock and eventually seal the natural permeability. This would decrease the leak-off at late time and could explain the conjoint observation of limited fracture extent and post-shut-in growth. Such a polymer filter cake buildup process is not accounted for in our model; further investigation would be required to confirm such an hypothesis.

% Also speak of potential thermal fractures taking up fluid ?

\begin{figure}[ht!]
    \centering
    \includegraphics[width=1\linewidth]{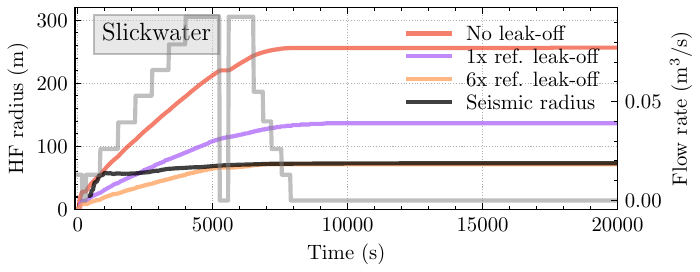}
    \caption{Accounting for fluid leak-off during the slick-water stage. To match the observed fracture radius we need to multiply the leak-off coefficient used for the cross-linked gel stage (see Fig. \ref{fig:newHF_stage3_addingLeakOff}) by a factor 6. This more elevated leak-off is explained by the lesser viscosity of the fracturing fluid.}
    \label{fig:newHF_stage2_addingLeakOff}
\end{figure}

We can do the same exercise of adjusting the leak-off to match the observed seismic radius for the slick-water stage, although here we don't have the same confidence in using the seismic radius as a proxy of a radial planar fracture radius. The associated results are shown in Figure \ref{fig:newHF_stage2_addingLeakOff}. We observe that in order to match the final seismic radius we need to multiply by 6 the leak-off compared to the value obtained in the cross-linked gel stage. This difference is explained by the difference in viscosity between the two fracturing fluids and would imply, under the hypothesis that the seismic radius can be used as a measure of the fracture radius here as well, that the upscaled permeability of the reservoir in this stimulation area would be around $1.8.10^{-15}$ m$^2$. This value appears elevated regarding what would be expected on average in the reservoir, if we choose to trust it, it could be explained as a consequence of the local presence of a few relatively conductive fractures taking up a lot of fluid. Another explanation is that the single hydraulic fracture assumption does not hold here and that several fractures are taking up fluid within the stimulated volume. The lesser resolution of the microseismic events localization does not allow us to confidently validate or invalidate this hypothesis.

To further grasp the amount of leak-off we need to introduce to match the observed seismic radius, we can compute the fracture efficiency of the modeled hydraulic fractures. The fracture efficiency $\mu$ is defined as $\mu = V_{hf} / V_{inj}$ ($V_{hf}$ the hydraulic fracture volume and $V_{inj}$ the fracturing fluid injected volume). The time series of the fracture efficiency for both the cross-linked gel stage and the slick-water stage simulations introduced above are given in Appendix \ref{Appendix:fracture_efficiency}. To give a single number, the fracture efficiency at the start of the ramp down in flow rate is about 60\% for the cross-linked gel and 10\% for the slick-water stage when adding enough leak-off to match the observed seismic radii. Again, this is more relevant for the cross-linked gel where we are more confident that the single planar fracture hypothesis holds than for the slick-water stage where the lesser seismic acquisition coverage does not allow us to confidently confirm this hypothesis, and do not allow us to confidently recover the evolution of the fracture radius with time.

\subsection{Modeling the stimulations as fluid-induced dilatant shear fractures}
\label{section:results:as_hydroshearing}

The other way to obtain hydraulic stimulation is via hydro-shearing; the development of a large fluid-induced dilatant shear fracture developing on a preexisting natural fracture/shear zone. As introduced in Section \ref{section:methods:shear_f_model}, under this hypothesis, we can have an idea of the injection pressure via the characteristic pressure $\Delta p^*$ defined in Eq. \ref{eq:theis_dpstar}. We can use it to evaluate if such a hydro-shearing hypothesis is realistic for both the stimulation stages. We evaluate the order of magnitude of $\Delta p^*$ for the two values of fracturing fluid viscosity (separated by 3 orders of magnitude) and the highest injection rate $Q_0 \simeq 0.1$ m$^3$/s, over a range of transmissibility increase due to shear-induced dilation. We take for the final hydraulic transmissibility the lower bound $kw_h = 10^{-13}$ m$^3$ and the upper bound $kw_h = 10^{-12}$ m$^3$. Under a simple model where the dilation coefficient goes linearly from a peak value $\tan(\psi_p)$ to zero over a critical slip distance $\delta_c$ (aligned with what is implemented in the numerical model defined previously) the final fracture opening is $u_n = \tan(\psi_p)\delta_c/2$. Under the cubic-law model, the associated transmissibility is $kw_h = u_n^3/12$. Taking a critical slip distance of $\delta_c = 2.5$ mm (aligned with the experimental results of \cite{Iyare2025}) these lower and upper bounds of transmissibility increase correspond to peak dilatancy coefficients of $0.85$ and $0.18$, respectively (or dilatancy angles of about $5^\circ$ and $10^\circ$). The resulting expected orders of magnitude of injection pressure are reported in Table \ref{tab:pressures_characteristic}. For the hydro-shearing hypothesis to be realistic, the pressure must remain bellow the confining stress, otherwise a hydraulic fracture would open. This first result hints that shear-induced dilation could accommodate the slick-water but not the thicker cross-linked gel where the predicted injection pressure is several order of magnitude higher than the confining stress, and hence tends to show that a tensile hydraulic fracture would eventually open.

\begin{table}
\fontsize{10}{12}\selectfont \caption{Characteristic injection pressures $\Delta p^* = Q_0 \mu / 4\pi kw_h$ assuming only shear-induced dilation. Compared to the expected normal effective stress $\sigma'_0$ expected to be between 14 and 22 MPa.}
\label{tab:pressures_characteristic}
\centering{}%
\begin{tabular}{p{2.4cm}cc}
\toprule 
 & $k w_h = 10^{-13}$ m$^3$ & $k w_h = 10^{-12}$ m$^3$ \tabularnewline
\midrule
slick-water \newline$\mu \simeq 10^{-4}$ Pa.s & $\simeq10$ MPa $\simeq \sigma'_0$ & $\simeq 1$ MPa $\ll \sigma'_0$\tabularnewline
cross-linked gel: \newline$\mu \simeq 0.1$ Pa.s & $\simeq 10^4$ MPa $\gg \sigma'_0$  & $\simeq 10^3$ MPa $\gg \sigma'_0$\tabularnewline

\bottomrule
\end{tabular}
\end{table}

We confirm this by running numerical simulations using the solver introduced in \ref{section:methods:pyfracx_model}. This time we use it in a mix-mode setting by having a non-zero shear component to the in-situ traction. We take the tractions associated with the closest orientation to shear failure given our preferred state of stress as defined in Figure \ref{fig:mohr_cirlce}. We run simulations for the two values of fracturing fluid viscosity and the two values of dilation defined before. For these simulations, where we suppose the shearing/reopening of an existing fracture we use the "shear zone cubic law" defined in Eq. \ref{eq:shear_zone_cubic_law} rather than the base cubic law. This allows us to set a higher storage coefficient that encapsulates both the compressibility of the fluid within the fracture/shear zone and the compressibility of the porous material within the preexisting fracture. The value of this storage coefficient greatly affects the radius of the shear fracture (a lesser storage coefficient leading to a larger slip front). We choose it here as $S = 5.10^{-10}$ Pa$^{-1}$. This value is in the range of expected compressibility of natural fractures in crystalline rocks, but it has been chosen so that the obtained slip front radii are coherent with the observed seismic radius. Changing its value would lead to radically different predicted shear fracture radii.

% TODO Need to ref Alexis's paper here

% TODO : the more elevated fluid storage under a NF reactoivation in opening for the cross-linked gel could be an alternative explanation to leakoff regarding the reduced farcture size 

All the parameters used in these simulations are reported in Table \ref{tab:material_properties_hydroshearing}. 

\begin{table}
\begin{threeparttable}
\fontsize{10}{12}\selectfont \caption{Material properties used for numerical simulations testing the validity of the hydroshearing hypothesis}
\label{tab:material_properties_hydroshearing}
\centering{}%
\begin{tabular}{ll}
\toprule 
Young's modulus $E$ & 42 GPa\tabularnewline
Poisson's ratio $\nu$\tnote{(*)} & 0.0\tabularnewline
Friction coefficient $f$\tnote{(2)} & 0.6\tabularnewline
Cohesion $C$\tnote{(2)} & $3$ MPa\tabularnewline
Peak dilatancy coefficient & 0.08 or 0.18\tabularnewline
Eff. normal stress $\sigma_0$\tnote{(3)} & 23.254 MPa\tabularnewline
Shear stress $\sigma_0$\tnote{(3)} & 9.25 MPa\tabularnewline
Water compressibility (both fluids)\tnote{(*)} & $6.10^{-10}$ Pa$^{-1}$ \tabularnewline
Viscosity $\mu$ (slick-water)\tnote{(*)} & $1.3.10^{-4}$ Pa.s\tabularnewline
Viscosity $\mu$ (cross-linked gel)\tnote{(*)} & 0.1 Pa.s\tabularnewline
Fracture storage coefficient $S$ & $5.10^{-10}$ Pa$^{-1}$\tabularnewline
Fracture initial transmissibility $T_0$\tnote{(4)} & $10^{-14}$ m$^{3}$\tabularnewline
\bottomrule
\end{tabular}
\begin{tablenotes}
  \small
  \item[*] See Table \ref{tab:material_properties}
  \item[2] Coherent with results of \cite{Iyare2025}
  \item[3] From our "preferred" state of stress
  \item[4] Aligned with values found in the literature, taken to be small compare to the post-stimulation transmissibility (either after hydraulic fracturing or hydro-shearing). % TODO need a ref here
\end{tablenotes}
\end{threeparttable}
\end{table}

\begin{figure}[ht!]
    \centering
    \includegraphics[width=1\linewidth]{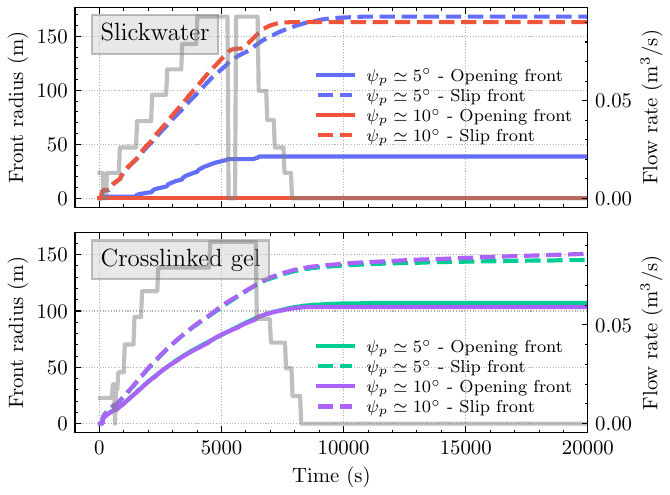}
    \caption{Shear and tensile fracture radii for both fracturing fluids and for two values of dilation angle. The parameters used are reported in Table \ref{tab:material_properties_hydroshearing}. We do not account for fluid leak-off here. The shear fracture radius is taken as the slip front, the tensile fracture radius is taken as the opening front after retrieving to it the shear-induced dilation.}
    \label{fig:hydroshearing_impermeable}
\end{figure}

The resulting fracture radius of these 4 simulations (2 values of fluid viscosity $\times$ 2 values of dilation coefficient) is shown in Figure \ref{fig:hydroshearing_impermeable}. We plot the slip front radius and the pure tensile opening front, which is obtained by subtracting to the total opening the opening due to shear-induced dilation. The results are aligned with what we could expect when looking at the order of magnitude of $\Delta p^*$: for the cross-linked gel a slipping patch develops but the transmissibility increase associated with dilation is not enough to accommodate the thick fluid, the pressure buildup eventually leads to the opening of an hydraulic fracture that accommodate most of the fluid. For the slick-water however, we observe that under the high dilatancy hypothesis the shear-induced dilation allows the fluid to remain below the confining normal stress and no hydraulic fracture develops For the lesser value of dilatancy, however, a hydraulic fracture does develop but its size remains small compared to the size of the slipping patch.

We also observe immediate arrest of the slipping patch upon shut-in for the slick-water stage, which is compatible with our observations derived from the microseismic activity. This indicates that, under our set of chosen parameters, the shear fracture propagates in the marginally pressurized regime.

\begin{figure}[ht!]
    \centering
    \includegraphics[width=1\linewidth]{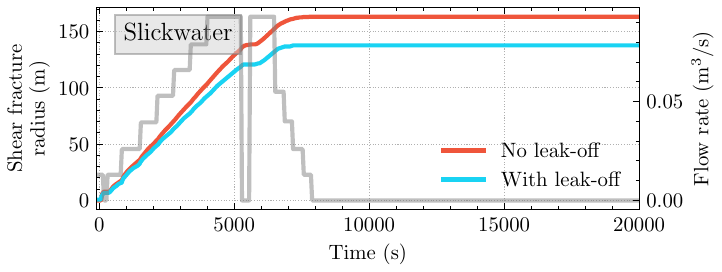}
    \caption{Effect of adding leak-off on the slip front radius for the slick-water stage under the elevated dilatancy hypothesis ($\psi_p \simeq 10^\circ$). Using the same leak-off coefficient value as used in the HF simulations to match the observed seismic radius (see Fig. \ref{fig:newHF_stage2_addingLeakOff}). The leak-off has a lesser effect under this fracture reactivation hypothesis due to the higher fracture storage than under the new HF hypothesis.}
    \label{fig:hydroshearing_effect_leakoff}
\end{figure}

We also experimented with adding leak-off to these "hydro-shearing" simulations. Due to the higher fluid storage within the fracture than under the pure HF hypothesis, the leak-off has less effect on the fracture size (of the order of 10\% both for the slip front and opening front). This is shown in Figure \ref{fig:hydroshearing_effect_leakoff} where we use the same leak-off parameters (bulk conductivity and storage) as needed to reproduce the observed seismic radius in the pure HF simulations shown in Figure \ref{fig:newHF_stage2_addingLeakOff}.

\subsection{
Influence of assumed cross-linked gel viscosity on the results
\label{section:incluence_viscosity}
}

Our modeling tools, both analytical and numerical, suppose a fluid with a Newtonian rheology of constant viscosity. Such a rheology is a good first-order approximation for the slick-water but less so for the cross-linked gel, which not only exhibits a non-Newtonian shear-thinning rheology but also see its polymer structure degraded when exposed to high temperatures, leading to a decrease in viscosity with time. We have assumed in the previous section that this strong simplification on the fluid rheology for the cross-linked gel, using a Newtonian rheology with a constant "effective" viscosity, still provides relevant first-order results. In this section, we run a sensitivity analysis on the different simulations results shown before, varying the value of this effective viscosity of the cross-linked gel. 
For that we consider varying this viscosity in the range $[0.025, 0.4]$ Pa.s, which correspond to one fourth to four time our previous base value of $0.1$ Pa.s. 

\begin{figure}[ht!]
    \centering
    \includegraphics[width=1\linewidth]{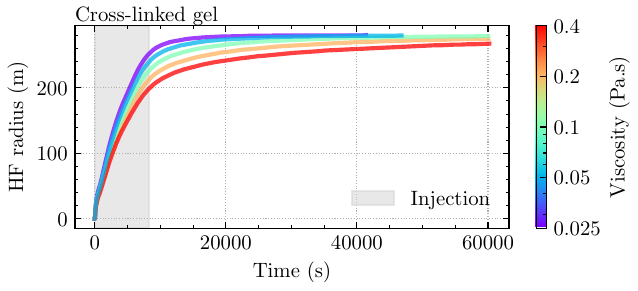}
    \caption{
    Evolution of the hydraulic fracture radius with time for the injection schedule of stage 3 considering a range of values of fluid viscosity to model the cross-linked gel stimulation stage.
    }
    \label{fig:viscosity_sens_newHF_noleakoff}
\end{figure}

We start by evaluating the robustness of the observation made in Section \ref{section:results:as_hf} that the hydraulic fracture keeps on propagating after shut-in. The results are shown in Fig. \ref{fig:viscosity_sens_newHF_noleakoff}. We observe that changing the value of viscosity leads to a different fracture radius at shut-in but does not affect the final fracture radius. This matches with the results of \cite{MoLe21} for the arrest of a hydraulic fracture propagating in the viscosity-dominated regime. 
% We recover the viscosity-dominated fracture radius scaling during the injection and obtain that the fracture radius at shut-in is proportional to the viscosity to the power $-1/9$. 
As such, the post-shut-in propagation amplitude and duration increases with the viscosity, but even for the lower bound of our range of $\mu=0.025$ Pa.s we still observe some post-shut-in growth of the fracture in the simulation results.

\begin{figure}[ht!]
    \centering
    \includegraphics[width=1\linewidth]{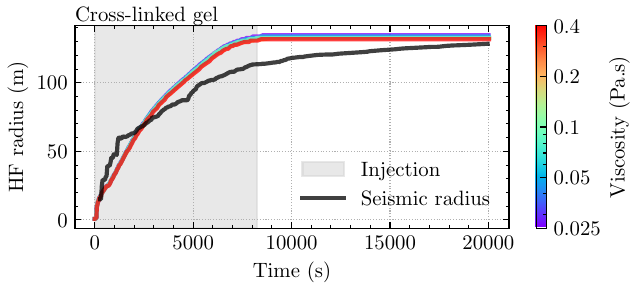}
    \caption{
    Evolution of the hydraulic fracture radius with time for the injection schedule the cross-linked gel stage, considering a range of values of fluid viscosity. Considering leak-off of fluid in the surrounding rock, but taking the leak-off parameters as independent of the fracturing fluid viscosity, as these parameters were chosen so that the simulated fracture extent matches the observations.
    }
    \label{fig:viscosity_sens_newHF_wleakoff}
\end{figure}

We also run our pure hydraulic fracture simulations with leak-off for various values of fracturing fluid viscosity. We choose to keep the same leak-off properties (permeability of the rock and viscosity of the fluid in the rock) as these were chosen so to match the observed fracture radius. The results are shown in Fig. \ref{fig:viscosity_sens_newHF_wleakoff}. We observe that when considering this amount of leak-off the viscosity has no effect on the hydraulic fracture, this is expected for a leak-off dominated HF if considering the Carter's leak-off coefficient to be independent of the fracturing fluid's viscosity.

\begin{figure}[ht!]
    \centering
    \includegraphics[width=1\linewidth]{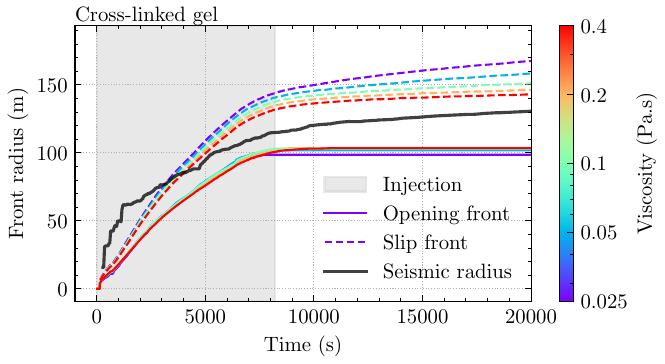}
    \caption{
    Shear and tensile fracture radii evolution considering a range of viscosity values for modeling the cross-linked gel stage. Here under the elevated value of dilation with $\psi_p \simeq 10^\circ$.
    }
    \label{fig:viscosity_sens_as_hydroshearing}
\end{figure}

Finally, we test the robustness of the observation made in Section \ref{section:results:as_hydroshearing} that when considering the cross-linked gel stage as the shearing/opening of a pre-existing fracture, realistic values of shear-induced dilation cannot accommodate this thick fluid and that a tensile hydraulic fracture eventually opens. We run the same simulation but considering the range of viscosity introduced above. The results are shown in Fig. \ref{fig:viscosity_sens_as_hydroshearing}. We observe that a smaller viscosity leads to a greater slip patch but has almost no effect on the size of the opening patch that develops behind it, even when considering an elevated value of shear-induced dilation $\psi_p\simeq 10^\circ$. This confirms that the cross-linked gel stimulation cannot be explained as a hydro-shearing process but only as a tensile hydraulic fracture, with possibly some shearing ahead of the tensile opening patch.

\subsection{Propagation of the uncertainty on the in-situ state of stress and the natural fractures orientation}
\label{section:results:uncertainty_propagation}

In Section \ref{section:data:state_stress} we have noted that the state of stress in the reservoir is far from being known exactly. 
In this section, we propose a workflow for propagating the uncertainty on the in-situ stress through our numerical simulations, both under the hypothesis of a  hydraulic fracture (HF) and a natural fracture (NF) reactivation (in shear or/and opening). In the second case, we also propagate the uncertainty on the natural fractures orientation.

We make use of the probabilistic state of stress model established in Section \ref{section:data:state_stress}:
\begin{itemize}
    \item the vertical stress gradient magnitude taken as reliable (or far less uncertain than its horizontal components counterparts), and equal to $\partial_z\sigma_v = 25.6$ kPa/m.
    \item the minimum horizontal stress magnitude within a range $\partial_z\sigma_h \in [15.2, 18.8]$ kPa/m.
    \item the maximum horizontal stress magnitude within a range of values that depends of the value of $\partial_z\sigma_h$, as represented in Figure \ref{fig:state_of_stress_model}.
    \item the pore pressure is taken as hydrostatic (and well known).
    \item the orientation of the horizontal stresses is centered on N$20^\circ$E, contained in the range N$5^\circ$E to N$30^\circ$E.
\end{itemize}

\begin{figure}[ht!]
    \centering
    \includegraphics[width=1\linewidth]{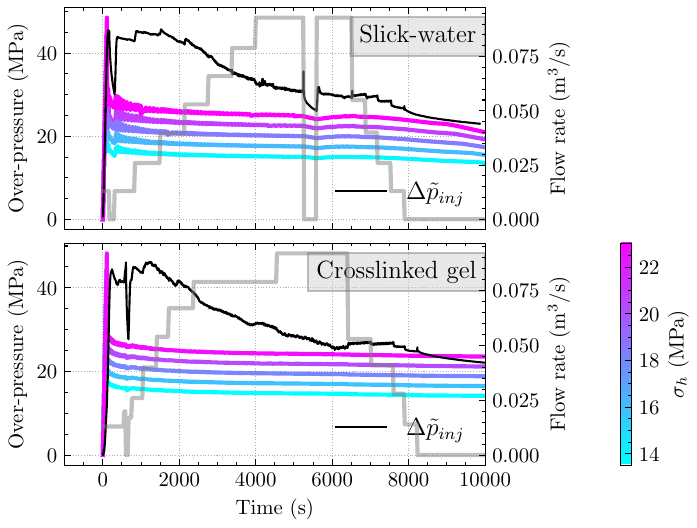}
    \caption{Injection over-pressure under the new HF hypothesis for 5 different values of confining stress spanning the range of expected values (see in-situ stress model in Fig. \ref{fig:state_of_stress_model}. Is also plotted $\Delta \tilde p_{inj}$ our estimate of the fracture injection over-pressure derived from the measured well-head pressure (see Section \ref{section:data:stimulation_data}). For both fracturing fluid viscosity values.}
    \label{fig:newHF_varShmin_pressure}
\end{figure}

We multiply the gradients by the respective true vertical depths of both stimulation stages: 2500 m for the cross-linked gel and 2556 m for the slick-water stage. For running simulations under a new HF hypothesis we start by taking uniformly 5 values of $\sigma_h$ over the defined range and retrieve the pore pressure to it to obtain the effective normal stress. For each of these values we run simulations identical to the one run in Section \ref{section:results:as_hf} (with leak-off), see Table \ref{tab:material_properties} for the value of the input parameters used. In Figure \ref{fig:newHF_varShmin_pressure} we plot the simulated injection pressure, compared to $\tilde p_{inj}$, the wellhead pressure to which we have removed a modeled friction losses part (see Section \ref{section:data:stimulation_data}). 
We observe that the observed injection pressure is outside the range of modeled pressures for the slick-water stage and in the very upper part of the range of modeled injection pressure for the slick-water stage. We also observe that for both stages we do not recover the decreasing trend in injection pressure. As proposed in \cite{McClure2023}, a possible explanation for this elevated pressure and the decreasing trend is the degradation/erosion of an initially elevated near-wellbore flow tortuosity in response to proppant placement. However proppant was only added to the slurry during the cross-linked gel, the associated proppant addition rate are given in Appendix \ref{Appendix:proppant}. No proppant was pumped for the slick-water stage. The addition of proppant can also cause a decrease in wellhead pressure by hydrostatic effects due to the higher density of the obtained slurry.
At late times, the measured pressure is close to the predicted pressures for the highest values of $\sigma_h$. This would indicate a locally elevated minimum horizontal stress under the assumption that, at late times, the near-wellbore tortuosity has been degraded.

\begin{figure*}[ht!]
    \centering
    \includegraphics[width=1\linewidth]{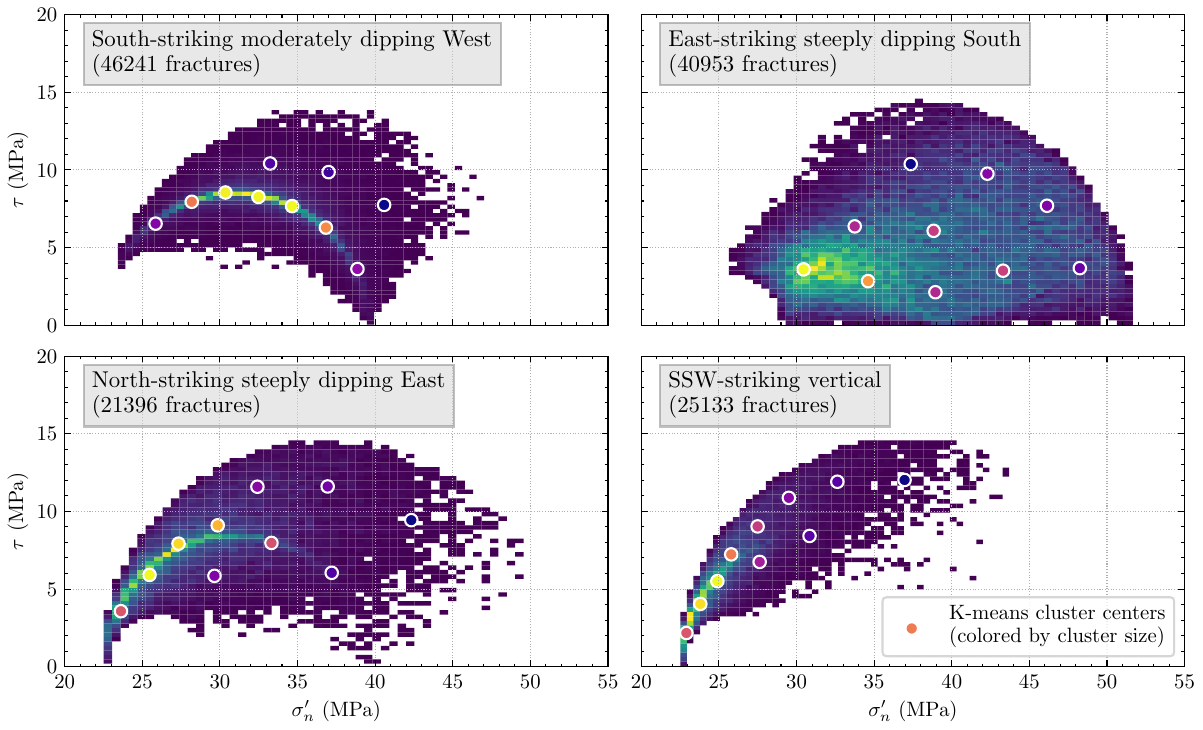}
    \caption{Using K-means to subsample the tractions distributions (shown as a 2D histogram) obtained by joint-sampling of the DFN orientation and the stress state. The cluster centers give the traction samples while the cluster sizes give an idea of the probability associated to these sample.}
    \label{fig:tractions_sampling}
\end{figure*}

For running simulations under the NF reactivation hypothesis we need the information of the full stress tensor combined with the orientation of the natural fractures (as our single planar fracture model requires uses projected traction values as input). For that we rely on the DFN realizations of \cite{Finnila2023}. We attribute to each fracture of these realizations the fracture set it is the closest to (out of the 4 included in the DFN model, see Figure \ref{fig:dfn_model_orienttaion}). Then, for each of the 5 values of $\sigma_h$ and for each of these 4 fracture sets, we draw $N$ values of the two remaining uncertain stress components: $\sigma_H$ and $\alpha_H$, $N$ being the number of fractures in the DFN realization associated with a given fracture set. 
We then project the obtained stress tensors onto the fractures orientation and obtain distributions of normal and shear stresses encompassing both the uncertainty on the stress model and on the natural fracture orientation. We use the K-means clustering algorithm as a way to efficiently  subsample these distributions. A more traditional Monte-Carlo method would require sample sizes too large to be numerically tractable. For each value of $\sigma_h$ and each fracture set, K-means is run on the $N$ 2D vectors of projected tractions. We obtain $M$ clusters, the clusters center are used as traction sample points that we pass to our solver while the cluster size can be used to have an idea of the prior density to attribute to each of these traction samples. Taking $M=10$ we end up with $M\times 4$ fracture sets $\times$ $5$ values of $\sigma_h$ $\times$ $2$ depths for the two stimulation stages, hence a total of $400$ tractions values. In Figure \ref{fig:tractions_sampling}, we show the traction distributions we obtain after this joint-sampling of the DFN natural fractures orientation and the state of stress for one value of $\sigma_h$ and how it is sub-sampled using K-means.

We run numerical simulations similar to the ones defined in \ref{section:results:as_hydroshearing} (with leak-off) for all these in-situ traction values, using the same values of input parameters as defined in Table \ref{tab:material_properties_hydroshearing} and for the smaller value of dilation coefficient ($\psi_p \simeq 5^\circ$) deemed more realistic regarding the results of \cite{Iyare2025}. We post-process the result by computing weighted percentiles / average using as weights the cluster sizes returned by K-means, this allows us to account in the final result visualization that some states of tractions are more likely than others. We represent the results by DFN fracture set.

\begin{figure}[ht!]
    \centering
    \includegraphics[width=1\linewidth]{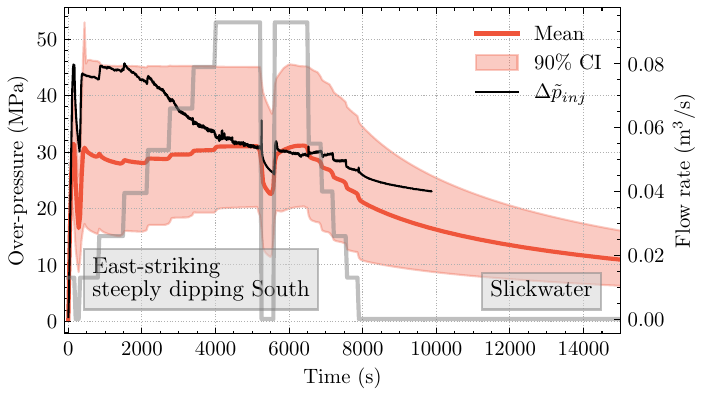}
    \caption{Propagating the DFN and the stress state uncertainty through our mix-mode numerical solver to obtain a confidence interval on the predicted injection pressure for one given DFN fracture set. Using the material parameters defined in Table \ref{tab:material_properties_hydroshearing}, for the elevated value of peak dilatancy $\psi_p=0.18$. $\Delta\tilde p_{inj}$ is our estimate of the injection over-pressure derived from the observed well-head pressure (see Section \ref{section:data:stimulation_data}).}
    \label{fig:NF_reactivation_pressure_stage2}
\end{figure}

For the slick-water stage, we observe that the reactivation (in shear) of fractures belonging to the fracture set "East-striking steeply dipping South", allows us to reproduce the observed elevated values of injection pressure within our confidence interval, as shown in Figure \ref{fig:NF_reactivation_pressure_stage2}. This set of simulations also grasp the pressure trend regarding the rate down with sharp pressure drops due to the higher gradients of pressure profile under such a fluid-induced shear fracture hypothesis than under a NF hypothesis. Nonetheless, the hypothesis that a fracture of this orientation would have accommodated the fluid cannot be verified looking at localized microseismic events and is not in agreement with the observed mean seismicity plane.

\begin{figure}[ht!]
    \centering
    \includegraphics[width=1\linewidth]{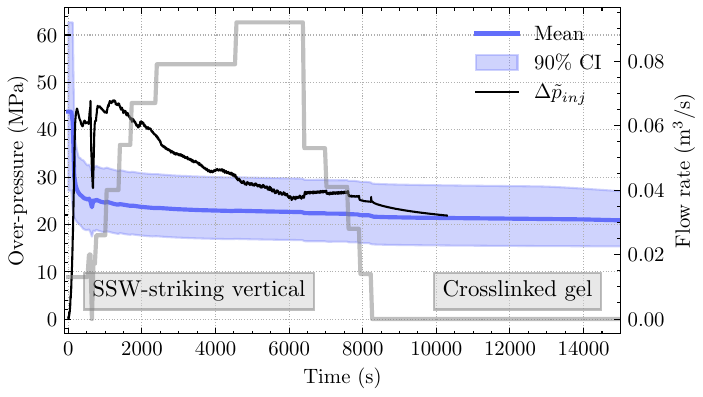}
    \caption{Propagating the DFN and the stress state uncertainty through our mix-mode numerical solver to obtain a confidence interval on the predicted injection pressure for one given DFN fracture set. Using the material parameters defined in Table \ref{tab:material_properties_hydroshearing}. $\Delta\tilde p_{inj}$ is our estimate of the injection over-pressure derived from the observed well-head pressure (see Section \ref{section:data:stimulation_data}).}
    \label{fig:NF_reactivation_pressure_stage3}
\end{figure}

For the cross-linked gel stage, we show the results we obtain for a fracture set "SSW-striking vertical", which would be relatively coherent with the observed seismicity plane, in Figure \ref{fig:NF_reactivation_pressure_stage3}. There, we show that the NF reactivation (in tensile opening) of a fracture belonging to this fracture set allows us to better match the observed pressure in late time than the new HF hypothesis, due to a slightly more elevated confining stress than when opening against $\sigma_h$ under the new HF hypothesis. 

\section{Conclusions}

In this work, we have investigated the nature of two hydraulic stimulation stages performed in April 2022 at the Utah FORGE EGS test site, differing primarily by the viscosity of the fracturing fluid used: a low-viscosity slick-water stage and a high-viscosity cross-linked gel stage. Both stages exhibited similar wellhead pressure responses but significantly different spatial and temporal distributions of induced microseismicity, particularly in their post-shut-in behavior. The cross-linked gel stage showed sustained microseismic activity for hours after shut-in, while the slick-water stage showed immediate cessation of microseismicity upon shut-in.

We first analyzed the microseismic data to extract information about the geometry and evolution of the stimulated volume. For the cross-linked gel stage, an adequate seismic acquisition coverage allowed us to confidently retrieve a time-dependent fracture radius from the located events, revealing that they are well distributed on a nearly vertical plane striking N15$^\circ$E. For the slick-water stage, the poorer seismic coverage resulted in higher location uncertainties and is potentially the reason of a more volumetric distribution of events, making the interpretation of a single planar fracture less certain.

Using scaling relationships on fracture size from analytical model of radial hydraulic fracture, we were able to explain the observation that when using cross-linked gel the fracture kept growing after shut-in while it immediately stopped when using slick-water. Using scalings from fluid-induced shear fracture theory we observed that an hydro-shearing hypothesis could be realistic for the slick-water stage but not for the cross-linked gel stage.

We then confirmed these insights using a 3D axisymmetric fully-coupled hydro-mechanical numerical model capable of resolving both tensile and shear failure, as well as fluid leak-off into the surrounding rock mass. The numerical simulations of HF assuming an impermeable host rock reproduced the key observations: post-shut-in growth for the cross-linked gel under a pure hydraulic fracture hypothesis, and immediate arrest for the slick-water stage. To match the cross-linked gel stage  observed fracture radius required accounting for significant but not unrealistic fluid leak-off. Numerical simulations under the shear fracture hypothesis showed that for the slick-water stage, shear-induced dilation could accommodate the fluid without opening a tensile fracture, provided the dilatancy is sufficiently high. However, for the cross-linked gel stage, the simulations confirmed that shear-induced dilation alone cannot accommodate the viscous gel, leading to the opening of a tensile hydraulic fracture.

Finally, we proposed a workflow to propagate the significant uncertainty on the in-situ state of stress, as well as the uncertainty on natural fracture orientations derived from discrete fracture network models, through our numerical simulations. This uncertainty propagation exercise revealed that the natural fracture reactivation hypothesis for the slick-water stage, specifically targeting the East-striking steeply South-dipping fracture set, could reproduce the observed elevated injection pressure, although this orientation is possibly inconsistent with the observed mean seismicity plane. For the cross-linked gel stage, the reactivation of fractures oriented close to the observed seismicity plane (SSW-striking vertical) in tensile opening mode provided a slightly better match to the late-time injection pressures compared to the new hydraulic fracture hypothesis, due to a slightly more elevated confining stress normal to this plane.

This study leaves several questions open for future investigation. A key unresolved issue for the cross-linked gel stage   is the apparent contradiction between the observed post-shut-in growth (reproduced in our simulations only for an   impermeable medium) and the limited fracture extent (indicating a significant effect of leak-off during propagation). Including the effect of a polymer cake buildup in our leak-off model could allow us to reconcile these observations. Additionally, it would be interesting to introduce better suited fluid rheological models, especially for the modeling the degradation of the cross-linked gels viscosity with high temperatures. Finally, applying this analytical-numerical workflow to other stimulation sites, especially those with complementary geodetic data (such as distributed strain sensing or tiltmeters), would be extremely valuable to further extend and improve the proposed methodology.

\paragraph{Author Contributions}
SB: Conceptualization, Methodology, Formal
analysis, Investigation, Software, Validation, Visualization, Writing—original draft. BL: Conceptualization, Methodology, Validation, Software, Supervision, Funding acquisition, Writing—review and editing.

\paragraph{Funding}
The results were obtained within the EMOD project (engineering model for hydraulic stimulation). The EMOD project benefited from a grant (research contract no. SI/502081- 01) and an exploration subsidy (contract no. MF- 021- GEO- ERK) of the Swiss federal office of energy for the Enhanced Geothermal System, geothermal project in Haute-Sorne, canton of Jura, which is acknowledged.

\bibliographystyle{apalike}
\bibliography{RMRE_forge}

\begin{appendices}
\section{
Evolution of cross-linked gel viscosity when facing thermal degradation
}
\label{Appendix:fcrosslinked_gel_viscosity}

\begin{figure}[!htbp]
    \centering
    \includegraphics[width=1\linewidth]{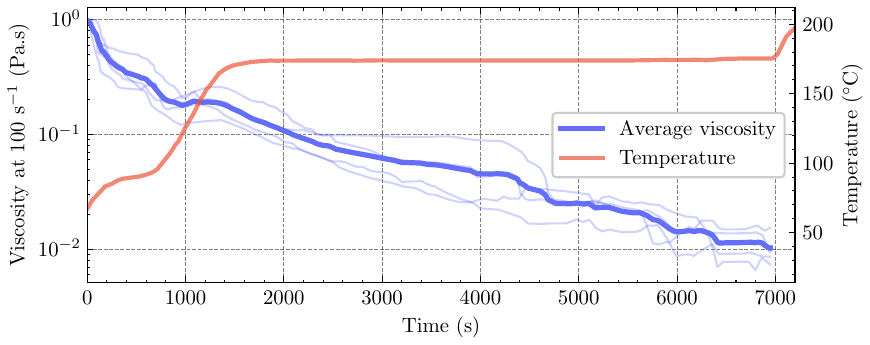}
    \caption{
    Evolution of the viscosity of the cross-linked CMHPG fracturing fluid when exposed to elevated temperatures of the order of the ones expected at the depths when the hydraulic stimulations are performed, data from \cite{UtahPhase3B}.
    }
    \label{fig:pressure_losses_stage2}
\end{figure}

\FloatBarrier

\section{Modeling pressure friction losses}
\label{Appendix:friction_losses}

\begin{figure}[!htbp]
    \centering
    \includegraphics[width=.8\linewidth]{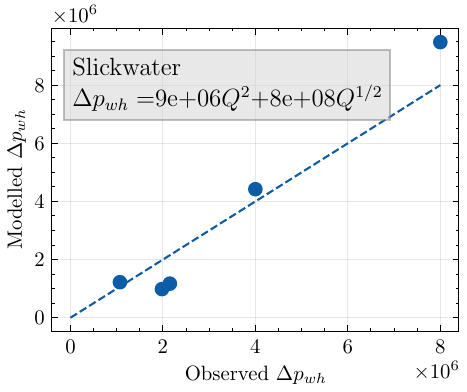}
    \caption{Observed immediate well-head pressure drop $\Delta p_{wh}$ vs modeled pressure drop $\alpha(Q_1^2 - Q_2^2) + \beta a(Q_1^{1/2} - Q_2^{1/2})$ for the slick-water stage.}
    \label{fig:pressure_losses_stage2}
\end{figure}

\begin{figure}[!htbp]
    \centering
    \includegraphics[width=.8\linewidth]{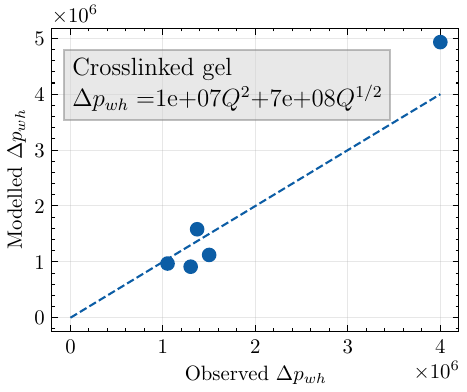}
    \caption{Observed immediate well-head pressure drop $\Delta p_{wh}$ vs modeled pressure drop $\alpha(Q_1^2 - Q_2^2) + \beta a(Q_1^{1/2} - Q_2^{1/2})$. For the cross-linked gel stage.}
    \label{fig:pressure_losses_stage3}
\end{figure}

\FloatBarrier % Requires \usepackage{placeins}

\section{
Sensitivity analysis of seismic radius estimation hyperparameters
}
\label{Appendix:sensitivity_analysis_seismic_radius}

\begin{figure}[!htbp]
    \centering
    \includegraphics[width=1\linewidth]{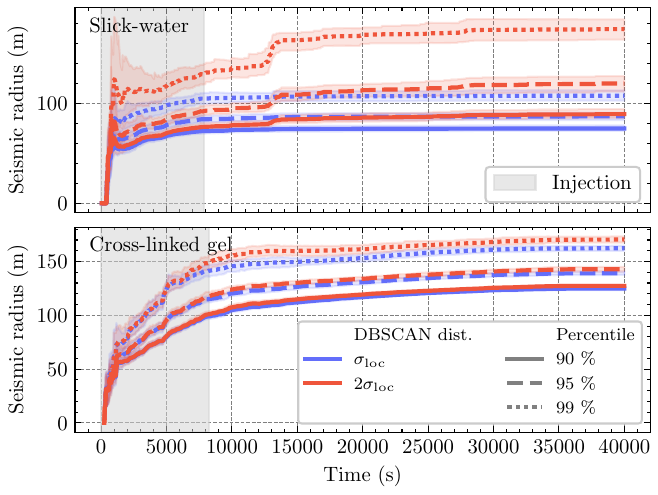}
    \caption{
    Sensitivity analysis of the seismic radius computation method regarding its two hyper-parameters: the DBSCAN neighborhood radius (or DBSCAN distance) and the percentile used when computing the seismic radius from a distribution of radial distances. $\sigma_{\text{loc}}$ is the median value of the location uncertainty of the events ($\sigma_{\text{loc}}\simeq 45.4$ m for the slick-water stage and $\simeq 34.6$ m for the cross-linked gel stage).
    }
    %for the no leak-off simulation
\label{fig:sensitivity_analysis_seismic_radius}
\end{figure}

The method introduced in Section \ref{section:data:seismic_radius} to compute from the located microseismic events a seismic radius evolving with time makes use of two hyperparameters both aiming at filtering out the events not co-located with the fluid-induced fracture: the neighborhood radius used in the DBSCAN algorithm, run once on all the events, and the percentile on the radial distances to the barycenter, used to compute the seismic radius at a given time. Figure \ref{fig:sensitivity_analysis_seismic_radius} shows how the computed seismic radius changes when we used different values of these two parameters. Looking at the results for the cross-linked gel stage, that benefited from a better seismic acquisition coverage, we observe that both parameters have a non negligible but not first-order effect on the results. This is expected given that, visually, the located microseismic events appear well co-located with an evolving radial fracture. For the slick-water stage however, the results are more dependent on these parameters. We notably observe jumps in the seismic radius for the more elevated DBSCAN neighborhood radius, caused by the inclusion of patches of microseismicity that are not spatially contiguous with the rest of the located microseismic activity.

\FloatBarrier

\section{Modeled hydraulic fracture treatment efficiencies}
\label{Appendix:fracture_efficiency}

\begin{figure}[!htbp]
    \centering
    \includegraphics[width=1\linewidth]{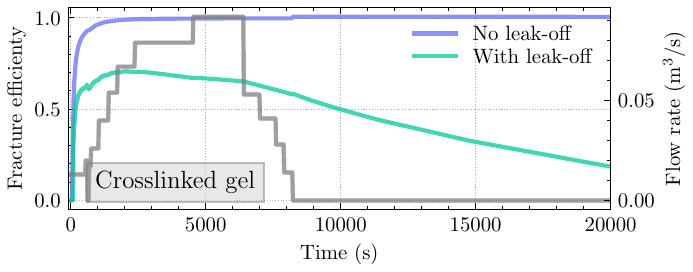}
    \caption{Modeled fracture efficiency for the cross-linked gel with and without leak-off (see results in Fig. \ref{fig:newHF_stage3_addingLeakOff}). The fracture efficiency is defined as $\mu = V_{hf} / V_{inj}$ ($V_{hf}$ the hydraulic fracture volume and $V_{inj}$ the fracturing fluid injected volume). We consider the volume injected at the well-head and consider line compressibility, this compressibility causes the efficiency to be lesser than one at the beginning (even for the zero leak-off case).} 
    %for the no leak-off simulation
\label{fig:stage3_modeled_fracture_efficiency}
\end{figure}

\begin{figure}[!htbp]
    \centering
    \includegraphics[width=1\linewidth]{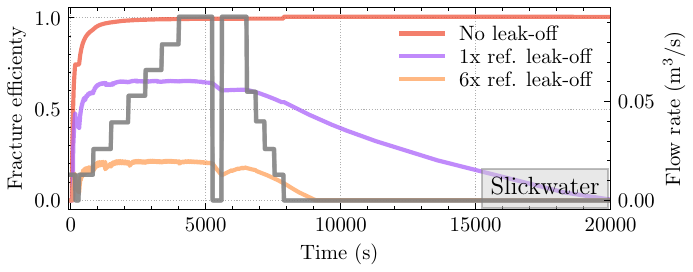}
    \caption{Same results as shown in Fig. \ref{fig:stage3_modeled_fracture_efficiency} but for the slick-water stage modeling results introduced in Fig. \ref{fig:newHF_stage2_addingLeakOff}.}
    \label{fig:stage2_modeled_fracture_efficiency}
\end{figure}

\FloatBarrier

\section{Proppant injection history during the cross-linked gel stage}
\label{Appendix:proppant}

\begin{figure}[!htbp]
    \centering
    \includegraphics[width=1\linewidth]{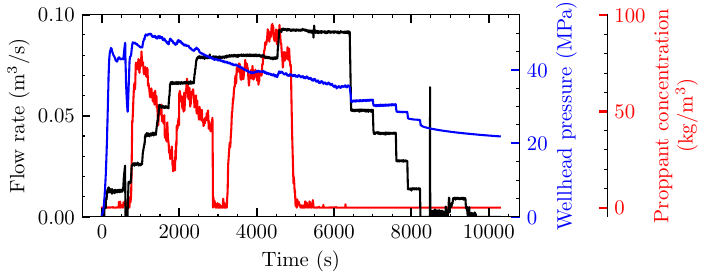}
    \caption{History of proppant injection during the cross-linked gel stage. The higher density of the slurry when proppant is added could partly explain the decreasing trend in well-head pressure that is not due to the hydraulic fracture mechanics. It is also likely due to a progressively degraded near wellbore tortuosity, which is the most likely explanation to explain this same decreasing trend in well-head pressure observed in the slick-water stage where no proppant was added.}
    \label{fig:stage3_proppant_history}
\end{figure}

\FloatBarrier

\section{Deviation from  the viscosity-storage scaling at early time for the cross-linked gel stage}
\label{Appendix:stage3_Mvertex_scaling}

\begin{figure}[!htbp]
    \centering
    \includegraphics[width=.8\linewidth]{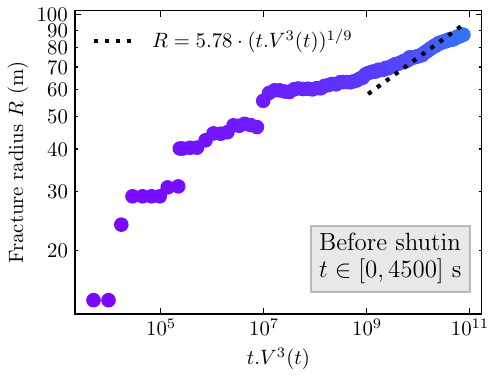}
    \caption{We do not exactly recover the scaling on the fracture radius for a viscosity-storage dominated regime, as determined by \citep{Savitski2002}, during the first half of the injection for the cross-linked gel stage. We do recover it for the second half of the injection and during the post-shut-in growth (see Fig. \ref{fig:recovering_mvertex_scaling}).}
    \label{fig:stage3_scaling_mvertex_from_t0}
\end{figure}

\FloatBarrier

\section{
Robustness of scaling recovery to seismic radius estimation hyperparameters
}
\label{Appendix:stage3_Mvertex_scaling_robustness}

\begin{figure*}[b]
    \centering
    \includegraphics[width=.8\linewidth]{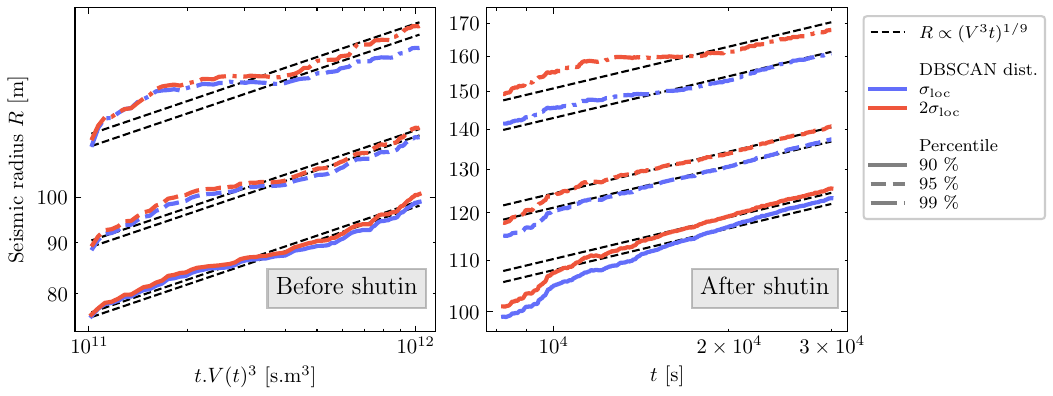}
    \caption{
    Sensitivity analysis of the results exposed in Fig. \ref{fig:recovering_mvertex_scaling} regarding the two hyper-parameters of the method used to estimate the seismic radius of the located microseismic events, used a proxy of the fracture radius. $\sigma_{\text{loc}}$ is the median value of the location uncertainty of the events.
    }
    \label{fig:recovering_mvertex_scaling_ensitivity_analysis}
\end{figure*}
\end{appendices}

\end{document}